\documentclass{article}
\usepackage{jheppub}
 \normalsize

\def\lsim{\raise0.3ex\hbox{$\;<$\kern-0.75em\raise-1.1ex\hbox{$\sim\;$}}}
\def\gsim{\raise0.3ex\hbox{$\;>$\kern-0.75em\raise-1.1ex\hbox{$\sim\;$}}}

\newcommand{\mat}[1]{\begin{pmatrix} #1 \end{pmatrix}}

\usepackage{graphics}
\usepackage{rotating}
\usepackage{multirow}
\usepackage{pstricks}
\usepackage{dcolumn}
\newcommand{\be}{\begin{eqnarray}}
\newcommand{\ee}{\end{eqnarray}}

\newcommand{\n}{\nonumber\\}

\def\bea{\begin{eqnarray}}
\def\eea{\end{eqnarray}}
\def\met{\slashed E_T}
\usepackage{epsfig}
\usepackage{slashed}
 \normalsize

\begin{document}
\title{Mono-jet, -photon and -$Z$ Signals of a Supersymmetric ($B-L$) model
 at the Large Hadron Collider}
\author{W. Abdallah$^{1,2}$, J. Fiaschi$^3$, S. Khalil$^1$ and S. Moretti$^{3}$ }
\vspace*{0.2cm}
\affiliation{$^1$Center for Fundamental Physics, Zewail City of Science and Technology, 6 October City, Giza, Egypt.\\
$^2$Department of Mathematics, Faculty of Science, Cairo
University, Giza, Egypt.\\
$^3$School of Physics and Astronomy, University of Southampton,
Highfield, Southampton, United Kingdom.}
\date{\today}

\abstract{
Search for invisible final states produced at the Large Hadron Collider (LHC) by new physics scenarios are normally carried out resorting to a variety of probes emerging from the initial state, in the form of single-jet, -photon and -$Z$ boson signatures. These are particularly effective for models of Supersymmetry (SUSY) in presence of $R$-parity conservation, owing to the presence in their spectra of a stable neutralino as dark matter candidate. We assume here as theoretical framework Supersymmetric ($B-L$) extension of the Standard Model (BLSSM), wherein a mediator for invisible decays can be $Z'$ boson. The peculiarity of the signal is thus that the final state objects carry a very large (transverse) missing energy, since the $Z'$ is naturally massive and constrained by direct searches and electro-weak precision tests to be at least in TeV scale region. Under these circumstances the efficiency in accessing the invisible final state and rejecting the standard model background is very high. This somehow compensates the rather meagre production rates. Another special feature of this invisible BLSSM signal is its composition, which is often dominated by sneutrino decays (alongside the more traditional neutrino and neutralino modes). Sensitivity of the CERN machine to these two features can therefore help disentangling the BLSSM from more popular SUSY models. We assess in this analysis the scope of the LHC in establishing the aforementioned invisible signals through a sophisticated signal-to-background simulation carried out in presence of parton shower, hadronisation and detector effects. We find that significant sensitivity exists already after 300~fb$^{-1}$ during Run 2. We find that mono-jet events can be readily accessible at the LHC, so as to enable one to claim a prompt discovery, while mono-photon and -$Z$ signals can be used as diagnostic tools of the underlying scenario.
}

\maketitle

\section{Introduction}
\label{sec:intro}

The minimal realisation of SUSY known as the Minimal Supersymmetric Standard Model (MSSM)
has come under increased pressure to explain current LHC data. On the one hand,
while still accommodating the existence of a Higgs boson compatible with the experimental measurements, the MSSM
suffers from severe fine-tuning in this respect, also known as the small hierarchy problem,
as the discovered Higgs boson mass of 125~GeV is 
dangerously close to its predicted absolute upper limit (130~GeV or so) in the MSSM (in fact, already
the LEP and Tevatron exclusion limits at around 115~GeV were posing such a problem), wherein the discovered
Higgs boson is identified
with the lightest CP-even Higgs state, whereas non-minimal versions of SUSY can place such a limit significantly higher, say,
below $2M_Z$. On the other hand, the total absence of SUSY signals, in the form of
particle state counterparts of the Standard Model (SM) objects (known as sparticles), rather than inspiring  the creation of contrived 
MSSM spectra to explain it, it should induce one to more naturally call for different SUSY cascade decays occurring in non-minimal versions of SUSY, owing to an additional neutralino entering as last decay step, thereby onsetting
decay topologies to which current SUSY searches are less sensitive than in the MSSM case.
 
In the light of all this, it has therefore become of relevance to explore non-minimal realizations of SUSY, better
compatible with current data than the MSSM yet similarly predictive and appealing theoretically. Because of the 
well established existence of non-zero neutrino masses, a well motivated path to follow in this direction is to consider the BLSSM. Herein, (heavy) right-handed neutrino
superfields are introduced in order to implement a type I
seesaw mechanism, which provides an elegant solution for the existence and
smallness of the (light) left-handed neutrino masses. Right-handed neutrinos can be
naturally implemented in the BLSSM, which is based on the gauge group $SU(3)_C \times
SU(2)_L \times U(1)_Y \times U(1)_{B-L}$, hence the
simplest generalisation of the SM gauge group (through an additional $U(1)_{B-L}$ symmetry). In this model, it has been shown that the scale of ($B-L$) symmetry breaking is related to the SUSY breaking
scale \cite{Khalil:2007dr}, so that this SUSY realization predicts several testable signals at the LHC,
not only in the sparticle domain but also in the $Z'$  (a $Z'$ boson in fact emerges from the $U(1)_{B-L}$
breaking), Higgs (an additional singlet state is economically introduced here, breaking the $U(1)_{B-L}$ group) and (s)neutrino sectors \cite{Khalil:2006yi,B-L-LHC,PublicPapers}. Furthermore, other than assuring its testability
at the LHC, in fact in a richer form than the MSSM (because of the additional (s)particle states), the BLSSM also alleviates the aforementioned
little hierarchy problem
of the MSSM, as both the additional singlet Higgs state and right-handed (s)neutrinos
\cite{BLMSSM-Higgs,O'Leary:2011yq,Basso:2012ew,Elsayed:2012ec,Khalil:2015naa}
release additional parameter space from the LEP, Tevatron and LHC constraints.
A Dark Matter (DM) candidate plausibly different from the MSSM one exists as well \cite{Basso:2012gz}.
Finally, interesting results on the ability of the BLSSM to emulate the Higgs boson signals isolated at the LHC Run 1
have also emerged, including the possibility of explaining possible anomalies hinting 
at a second Higgs peak in the CMS sample \cite{Abdallah:2014fra}.

While the BLSSM clearly represents an appealing framework for non-minimal SUSY, both theoretically and
experimentally, so as to deserve the phenomenological attention that the papers referred to above now exemplify,
it remains crucial to find a way of disentangling its experimental manifestations from those of
other non-minimal SUSY realizations. In this connection, it is obvious to mention that SUSY cascade decays may appear
rather similar in any non-minimal SUSY, as there are essentially no handles to identify the nature of the additional neutralino providing the last step of the new SUSY ladder,  the invisible (transverse) energy. Also, it is conceivable to expect that
the Higgs sectors of such non-minimal SUSY versions may be very difficult to extricate one from the other. In fact,
no matter the number and nature of additional Higgs states above and beyond the MSSM ones, the patterns of signals emerging are more often than not rather similar in all such non-minimal SUSY conceptions. This thus leaves the 
$Z'$ and (s)neutrino
sectors as ideal hallmark manifestations of the BLSSM as candidate underlying SUSY model. However, if one investigates
separately the $Z'$ and (s)neutrino dynamics, there is again little in the way of disentangling the BLSSM $Z'$ from that of  popular extended gauge models (with and without SUSY) or distinguishing the BLSSM (s)neutrinos from those of other SUSY
scenarios  (minimal or not). 

It may be different though if $Z'$ and (s)neutrino dynamics (of the BLSSM) are somehow tested together. In this respect,
from a phenomenological point of view, an intriguing signal, both for experimental cleanliness and theoretical naturalness, 
would be the one involving totally invisible decays of a $Z'$ into (s)neutrinos, thereby accessible in mono-jet, single-photon and $Z$-ISR
(Initial State Radiation) analyses (see \cite{mono-review} and references therein for a snapshot of the current LHC status
of the latter)\footnote{We do not consider here the case of mono-top and $W$-ISR probes, which have also been used experimentally.}. Contrary to SUSY models which do not have a $Z'$ in their spectra or where the invisible final state is induced
by direct couplings of the lightest neutralino pair to (light and potentially highly off-shell) $Z$ bosons, in the BLSSM one
can afford resonant $Z'$ production and decay into heavy (s)neutrinos which can in turn decay, again on-shell, into an invisible final state. Under these circumstances, one would expect the typical distributions of the visible probe
(whether it be mono-jet, single-photon or $Z$-ISR) to be substantially different from the case of other SUSY
scenarios. This remains true even if the $Z'$ decays into gauginos, either directly into the lightest neutralinos or else into heavier --ino states cascading down (invisibly) to the Lightest Supersymmetry Particle (LSP) and even 
(both light and heavy) neutrinos.

It is the purpose of this paper to systematically study this phenomenology and assess the scope of the LHC
in either constraining this scenario or accessing it. Our plan is as follows. In Sec.~\ref{sec:Zp+SnuBLSSM} 
we describe the theoretical setup of the $Z'$ and (s)neutrino sectors of the BLSSM.
In Sec.~\ref{sec:limits} we discuss experimental constraints on the model, from both Electro-Weak Precision Tests (EWPTs)
and direct searches.
The subsequent three sections are devoted to study the phenomenology of the three aforementioned
experimental probes, i.e.,  mono-jet, single-photon or $Z$-ISR, respectively. Finally, we
summarize and conclude in Sec.~\ref{sec:summa}. (A briefer account of the upcoming work has been given in Ref.~\cite{Abdallah:2015hma}.)

\section{$Z'$ and right-handed sneutrinos in the BLSSM}
\label{sec:Zp+SnuBLSSM}

In  the BLSSM, the particle content includes the
following fields in addition to the MSSM ones: three chiral
right-handed superfields ($\hat{N}_i$), a vector superfield associated
to $U(1)_{B-L}$ ($\hat{Z}'$) and two chiral SM singlet Higgs
superfields ($\hat{\chi}_1$, $\hat{\chi}_2$). The superpotential of this model
is given by
\bea {W} &=&
(Y_U)_{ij}\hat{Q}_i\hat{H}_u \hat{U}^c_j+(Y_D)_{ij}\hat{Q}_i\hat{H}_d\hat{D}^c_j+(Y_L)_{ij}\hat{L}_i\hat{H}_d\hat{E}^c_j
+ (Y_{\nu})_{ij}\hat{L}_i\hat{H}_u\hat{N}^c_j \nonumber\\
&+&(Y_N)_{ij}\hat{N}^c_i\hat{\chi}_1\hat{N}^c_j%
+ \mu(\hat{H}_u\hat{H}_d)+\mu'\hat{\chi}_1\hat{\chi}_2.
\label{super-potential-b-l}
\eea%
The ($B-L$) charges of superfields appearing in the superpotential
$W$ are given in Tab.~\ref{ub-l-charge}.
 \begin{table}[!t]
\begin{center}
{\small\fontsize{8}{8}\selectfont{
\bgroup
\def\arraystretch{1.5}
\begin{tabular}{|c|c|c|c|c|c|c|c|c|c|c|} \hline
 & $\hat{L}_i$ & $\hat{N}^c_i$ & $\hat{E}^c_i$ & $\hat{Q}_i$ & $\hat{U}^c_i$ & $\hat{D}^c_i$ & $\hat{H}_u$ & $\hat{H}_d$ & $\hat{\chi}_1$  & $\hat{\chi}_2$
  \\ \hline
{ $SU(2)_L\times U(1)_Y$}
 & $ ({\bf 2}, -\frac{1}{2})$  &  $({\bf 1}, 0)$  &  $({\bf 1}, 1)$
 & $({\bf 2}, \frac{1}{6})$&  $({\bf 1},-\frac{2}{3})$
 &  $({\bf 1}, \frac{1}{3})$& $({\bf 2}, \frac{1}{2})$& $({\bf 2}, -\frac{1}{2})$ &
$({\bf 1}, 0)$ &  $({\bf 1}, 0)$
  \\ \hline
 $U(1)_{B-L}$ & $-\frac{1}{2}$  &  $\frac{1}{2}$  &  $\frac{1}{2}$
 & $\frac{1}{6}$&  $-\frac{1}{6}$&  $-\frac{1}{6}$& 0& 0 & -1 & 1
   \\ \hline
\end{tabular}
\egroup
}}
\caption{\label{ub-l-charge} The $U(1)_{B-L}$ charges of the superfields in the BLSSM.}
\end{center}
\end{table}

For universal soft SUSY-breaking terms at the scale of Grand Unification Theories (GUTs),
$M_X$, the soft breaking Lagrangian is given by %
\bea %
-{\cal L}_{soft} &=& m^2_0\left[|\widetilde Q_i|^2+|\widetilde
U_i|^2+|\widetilde D_i|^2+|\widetilde L_i|^2+|\widetilde E_i|^2
\right.
+ \left.|\widetilde N_i|^2 +|H_u|^2+| H_d|^2+|
\chi_1|^2+|\chi_2|^2\right] \nonumber\\
&+& A_0\left[Y_{U}{\widetilde Q}H_u{\widetilde
U}^c+Y_{D}{\widetilde Q}H_d{\widetilde D}^c+Y_{E}{\widetilde
L}H_d{\widetilde E}^c + Y_{\nu}{\widetilde L}H_u{\widetilde
N}^c+Y_{N}{\widetilde N}^c\chi_1{\widetilde N}^c \right] \nonumber\\&+&
\left[B(\mu H_uH_d+\mu'\chi_1\chi_2)+h.c.\right]
+ \frac{1}{2}M_{1/2}\left[{\widetilde g}^a{\widetilde
g}^a+{\widetilde W}^a{\widetilde W}^a+{\widetilde B}{\widetilde
B}+{\widetilde Z'}{\widetilde Z'}+h.c.\right],~~~~
\eea %
where the tilde denotes the scalar components of the chiral matter
superfields and fermionic components of the vector superfields.
The scalar components of the Higgs superfields $\hat{H}_{u,d}$ and
$\hat{\chi}_{1,2}$ are denoted as  $H_{u,d}$ and $\chi_{1,2}$,
respectively.

As shown in Ref.~\cite{Khalil:2007dr}, both the ($B-L$) and
EW symmetry can be broken radiatively in supersymmetric theories. 
In this class of models, the EW, ($B-L$) and soft SUSY breakings can occur at the TeV scale. 
The conditions for EW Symmetry Breaking (EWSB) are given by %
\bea
\mu^2&=&\frac{m^2_{H_d}-m^2_{H_u}\tan^2\beta}{\tan^2\beta-1}-M^2_Z/2,~~~~~ 
\sin2\beta=\frac{2m^2_3}{m^2_1+m^2_2}\label{sm-higgs-condition},
\eea
 where
 \bea
&& m^2_i=m^2_0+\mu^2,~~i=1,2,~~~~~ m^2_3=-B\mu,~~~~~ \tan\beta=\frac{v_u}{v_d}, \nonumber\\&&
<H_u>=v_u/\sqrt2,~~~<H_d>=v_d/\sqrt2.
 \eea
Here $m_{H_u}$ and $m_{H_d}$ are the masses of the Higgs fields coupling to $u$ and $d$-type fermions,
respectively, defined at the
EW scale. Further, $M_Z$ is the mass of the neutral massive gauge boson in the SM. It is worth
noting that the breaking of $SU(2)_L\times U(1)_Y$ occurs at the
correct scale of the SM charged gauge boson mass ($M_W\sim80$ GeV).
Similarly, the conditions for the ($B-L$) radiative symmetry
breaking are given by
\bea
\mu'^{2}&=&\frac{\mu^2_{1}-\mu^2_{2}\tan^2\beta'}{\tan^2\beta'-1}-M^2_{Z'}/2,~~~~~
\sin2\beta'=\frac{2\mu^2_3}{\mu^2_1+\mu^2_2},\label{b-l-higgs-condition}
\eea
 where
 \bea
&& \mu^2_i=m^2_0+\mu'^{2},~~i=1,2,~~~~~\mu^2_3=-B\mu',~~~~~\tan\beta'=\frac{v'_1}{v'_2},\nonumber\\&&
<\chi_1>=v'_1/\sqrt2,~~~<\chi_2>=v'_2/\sqrt2.
\eea
  Here $m_{\chi_1}$ and $m_{\chi_2}$ are the $U(1)_{B-L}$-like Higgs masses at the TeV scale.
The key point for implementing radiative ($B-L$) symmetry
breaking is that the scalar potential for $\chi_1$ and $\chi_2$
receives substantial radiative corrections. In particular, a
negative squared mass would trigger  ($B-L$) symmetry breaking of
$U(1)_{B-L}$. After ($B-L$)
symmetry breaking has taken place, the $U(1)_{B-L}$ gauge boson acquires a
mass~\cite{Khalil:2006yi}: $M^2_{Z'}=g^2_{B-L}v'^2$. The experimental searches at high energy 
as well as precision measurements at lower scale 
impose    bounds on this mass. The most stringent constraint on the
$U(1)_{B-L}$ gauge boson mass is
obtained from LEP2 results, which imply
$ \frac{M_{Z'}}{g_{B-L}}>6$~TeV \cite{Cacciapaglia:2006pk}.
However, one should note that this bound is based on the assumption that  the
$Z'$ dominantly decays to SM quarks and leptons. If the $Z'$ decays to, e.g., right-handed (s)neutrinos 
with significant Branching Ratios (BRs), this bound is relaxed and much lighter $Z'$ masses are allowed \cite{Abdelalim:2014cxa}.  
This is indeed the BLSSM configuration that would at the same time favor searches for these decays in invisible final states,
that we are intending to tackle here.

We now consider the right-handed sneutrino sector in the BLSSM model. With a TeV scale
right-handed sneutrino, the sneutrino mass matrix, for one generation, in the basis ($\tilde{\nu}_L,\tilde{\nu}_L^\ast,\tilde{\nu}_R,\tilde{\nu}_R^\ast$), is given by the following $4\times 4$ Hermitian matrix:%
\begin{equation}
{\cal M}^2 = \mat{M^2_{LL} & M^2_{LR} \\ \left(M^2_{LR}\right)^{\dag} & M^2_{RR}
},\label{sneutrino-matrix}
\end{equation}
where
{\fontsize{10}{10}\selectfont
\begin{eqnarray}
M^2_{LL} & = & \left(m_{\tilde{L}}^2+m_D^2+\frac{1}{2}M_Z^2\cos 2\beta-\frac{1}{2}M_{Z'}^2\cos 2\beta'\right)\,\mathbf{1}_{2\times2},\\\n
M^2_{LR} & = & m_D(A_\nu-\mu\cot\beta + M_N)\,\mathbf{1}_{2\times2},\\\n
M^2_{RR} & = & \!\!\!\!\mat{M_N^2 + m_{\tilde{N}}^2 + m_D^2+ \frac{1}{2}M_{Z'}^2\cos 2\beta' & M_N(A_N-\mu'\cot\beta')\\
M_N(A_N-\mu'\cot\beta') & M_N^2+m_{\tilde{N}}^2+m_D^2+\frac{1}{2}M_{Z'}^2\cos 2\beta'},
\end{eqnarray}}
\noindent
where $m_D=Y_\nu v_u,\ M_N=Y_N v'_1$. It is clear that the mixing between left- and right-handed sneutrinos is quite suppressed since it is proportional to the Yukawa coupling $Y_\nu \lsim {\cal O}(10^{-6})$. Conversely, a large mixing between the right-handed sneutrinos and right-handed anti-sneutrinos is quite plausible, since it is given in terms of the Yukawa term $Y_N \sim {\cal O}(1)$.

\begin{figure}[h]
\begin{center}
\includegraphics[width=0.33\textwidth]{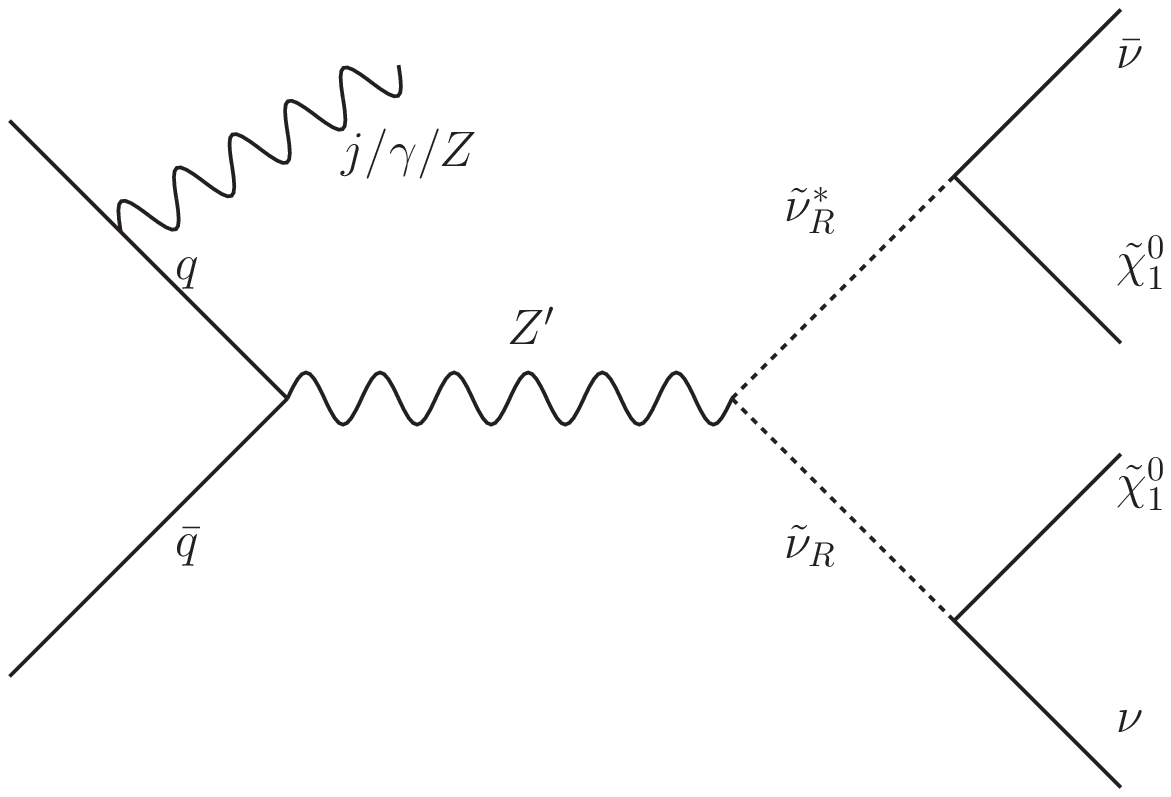}~\includegraphics[width=0.33\textwidth]{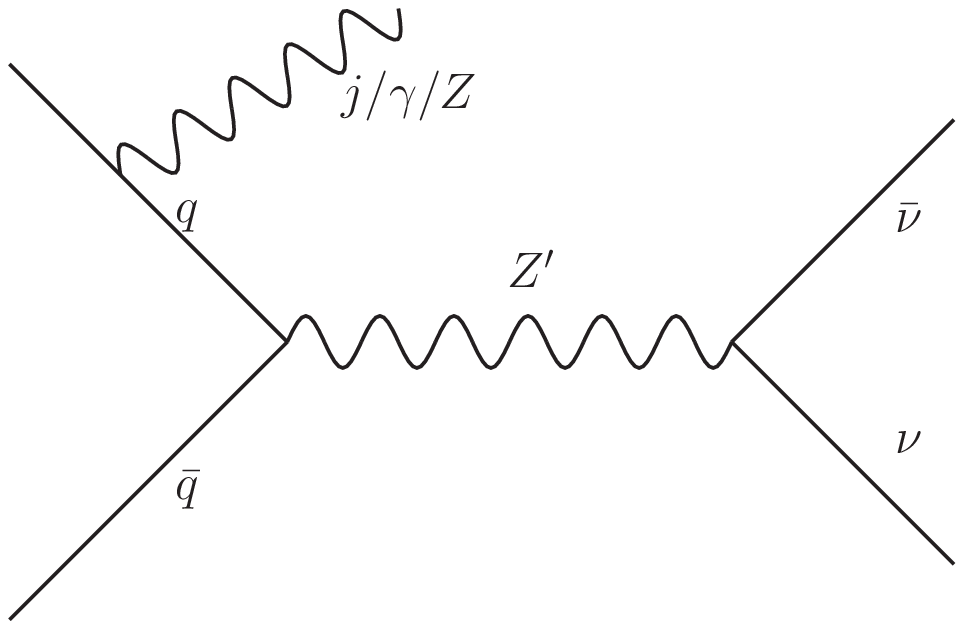}~\includegraphics[width=0.33\textwidth]{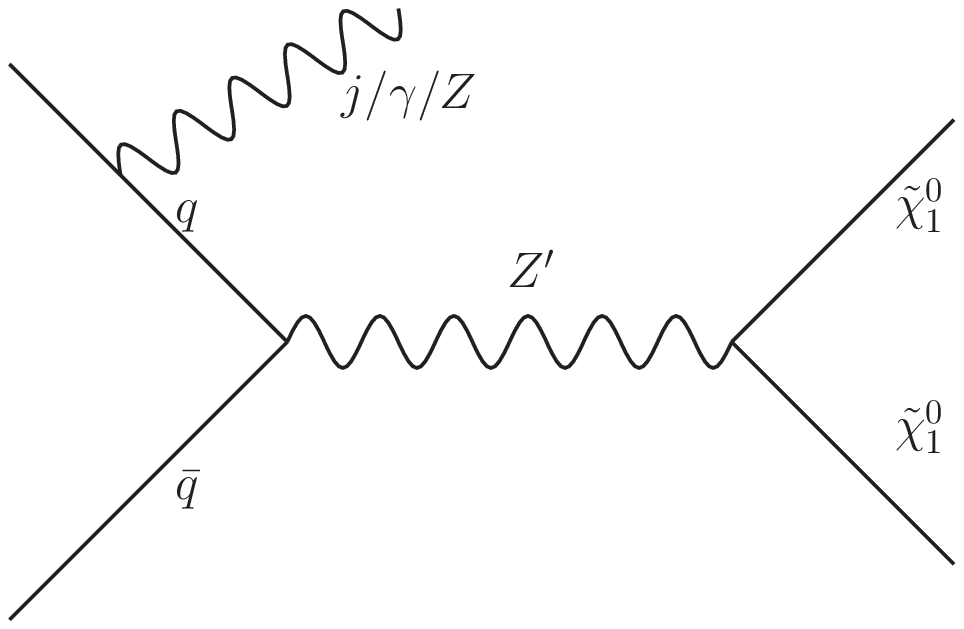}
\caption{Dominant Feynman diagrams contributing to the mono-jet, -photon and -$Z$ processes. (Notice that we neglect contributions to mono-$Z$ searches wherein the $Z$ is emitted from the final state.)}
\label{fig:mono_feynman}
\end{center}
\end{figure}

From the BLSSM Lagrangian, one can show that the relevant
interactions for the right-handed sneutrino are given by
\begin{eqnarray}
{\cal L}_{int}^{^{\tilde{\nu}_R}} &=& (Y_\nu)_{ij} \bar{l}_i P_R (V_{k2} \tilde{\chi}^+_k)^\dag (\Gamma_{\nu_R})_{\alpha j} \tilde{\nu}_{R_\alpha} + (Y_{\nu})_{ij}(U_{\rm MNS})_{il} \bar{\nu}_l P_R (N_{k1}^* \tilde{\chi}_k^0) (\Gamma_{\nu_R})_{j\alpha} \tilde{\nu}_{R_\alpha}  \nonumber\\
&+& (Y_\nu)_{ij} (M_N)_j \cos \beta \left[(\Gamma_{L_L})_{\beta i} \tilde{l}_\beta H^+ (\Gamma_{\nu_R})_{\alpha j}\tilde{\nu}_{R_\alpha} \right].
\end{eqnarray}
Here, we assume that the charged leptons are in their physical basis. The rotational matrices $\Gamma_{L}$ and $\Gamma_{\nu}$ are defined as $\Gamma_L \equiv (\Gamma_{L_L}, \Gamma_{L_R})$ and $\Gamma_\nu \equiv (\Gamma_{\nu_L}, \Gamma_{\nu_R})$. Further, the
neutralino mass matrix is diagonalized by a $4 \times 4$ rotation matrix $N$ and the chargino mass matrix is diagonalized by two rotation matrices $U,V$. From this, it can easily be concluded that, if the lightest right-handed sneutrino is lighter than the lightest slepton and lightest chargino, then it decays into light SM-like neutrinos and lightest neutralinos. This decay channel would be an invisible channel, since both light neutrinos and lightest
neutralinos would be escaping the detector. Hence, given the discussed SUSY construction peculiar to the BLSSM,
it can provide  a robust signature for BLSSM $Z'$ to sneutrino transitions through the
 mono-jet, single-photon  and $Z$-ISR  topologies that will be elaborated upon in the following sections. Competing invisible signals, though smaller in comparison, are direct $Z'$ decays into light neutrinos and lightest neutralinos, all other modes being essentially negligible over the
BLSSM parameter space we investigate (e.g., the $Z'$ could decay invisibly also via heavy gauginos, however, this dynamics is not specific to the BLSSM, so that we do not dwell on it here). 
Hence, the  Feynman diagrams relevant for our study are found in Fig.~\ref{fig:mono_feynman}.

\section{Experimental limits on the BLSSM}
\label{sec:limits}
In testing the extra neutral gauge sector of the BLSSM, we shall consider the experimental bounds coming from $Z'$ direct searches performed at the LHC Run 1  as well as limits stemming from EWPTs. On the one hand, 
indirect constraints have been widely reviewed in the literature and the most stringent limit has been computed in the context of a 
non-supersymmetric ($B-L$) model \cite{Cacciapaglia:2006pk}\footnote{Yet it is applicable here as we will set the mass scales of all sparticles to be much larger than
the EW scale.} and gives us an upper bound on the ratio between the $Z'$ mass and its gauge coupling at $99\%$ Confidence Level (CL):
\begin{equation}
M_{Z'} / g_{B-L} \gsim 6~\text{TeV}.
\end{equation}
On the other hand, direct $Z^\prime$ searches in Drell-Yan (DY) production have recently produced new bounds for heavy neutral resonances \cite{Khachatryan:2014fba,Aad:2014cka}: the exclusion limits that have been found from the LHC 8~TeV run, roughly speaking, forbid $Z'$ resonances with mass below 2~TeV.

However, these LHC  limits have been produced under the assumption of  a narrow $Z'$ resonance (i.e., $\Gamma_{Z^\prime} / M_{Z^\prime} \lesssim 10 \%$), condition which is specific only
to some  classes of models (e.g., $E_6$ motivated, Minimal $Z^\prime$, extra $(Z^*,W^*)$ doublet). Hence, they may not be applicable to the BLSSM, because
the large parameter space of such a SUSY model either grants one $Z'$ couplings to the fermions that can be small enough or 
affords one with more $Z'$ decay channels (or both) so as to keep the resonance hidden even below the declared threshold, since it has too low a cross section or is too broad (or both).

In order to illustrate this, we will propose a few benchmarks in both such scenarios.  With the exception of the limit extraction which has been done at Next-to-Next-to-Leading Order (NNLO) in QCD,  in all other cases, notice that that we have obtained our results in LO\footnote{Hence, the invariant mass
of the dilepton pair $M_{ll}$ and the Centre-of-Mass (CM) energy at the parton level $\sqrt{\hat s}$ are interchangeable quantities.}, 
which is generally a sufficient approximation in order to obtain our conclusions. Hereafter, we refer to the SM Background ($B$) as the subprocess $pp\to \gamma,Z\to l^+l^-$ $ (l=e,\mu)$ whereas the
$Z'$ Signal ($S$) is identified as the difference between the yield of the subprocess $pp\to \gamma,Z,Z'\to l^+l^-$ and that of the previous one, so that our significance $\alpha$ is given as
\bea
\alpha&=&2(\sqrt{S+B}-\sqrt{B}),~\text{for Poisson statistics},\\
\alpha&=&\frac{S}{\sqrt{S+B}},~\text{for Gaussian statistics},
\label{significance}
\eea
where $S$ and $B$ are given in terms of the number of events after 20(300) fb$^{-1}$ of integrated luminosity at the LHC Run 1(2).

\begin{figure}[t]
\begin{center}
\includegraphics[width=0.8\textwidth]{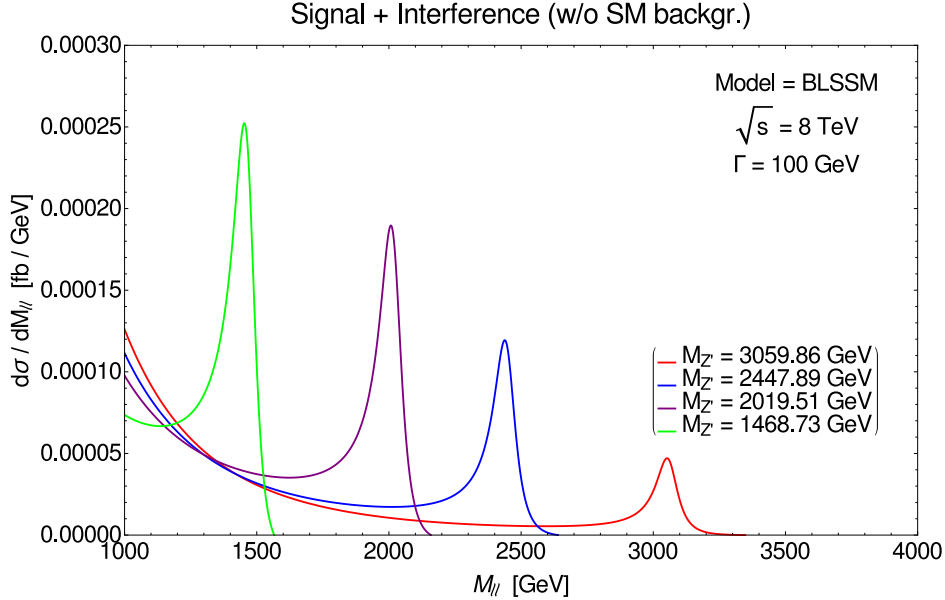}
\caption{Differential cross section distributions at LO in DY for the four benchmarks in Tab.~\ref{table:narrow}.}
\label{fig:Narrow_XS}
\end{center}
\end{figure}

\begin{figure}[t]
\begin{center}
\includegraphics[width=0.8\textwidth]{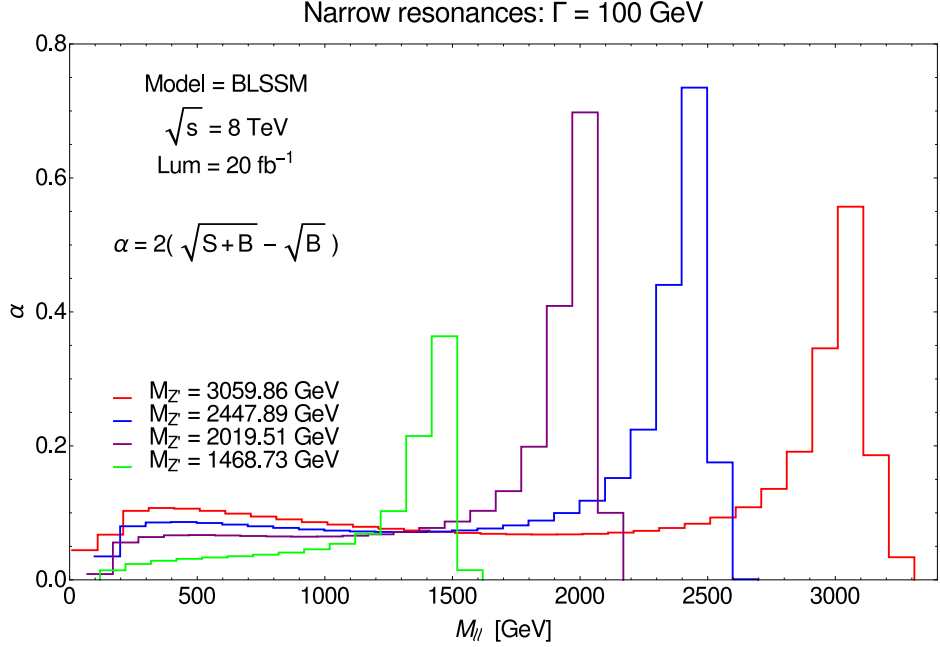}
\caption{Significance ($\alpha$) of the $Z^\prime$ signal in the dilepton channel for the four benchmarks in Tab.~\ref{table:narrow}. Poisson statistics has been assumed.}
\label{fig:Narrow_Sig}
\end{center}
\end{figure}

To start with, we put forward 
four benchmarks (see Tab.~\ref{table:narrow}) where the new gauge sector physics satisfies the EWPT constraints and at the same time the $Z^\prime$ boson would
have escaped the direct detection analysis of Run 1 data at the LHC, owing to small gauge couplings and/or a large width -- and consequent large interference effects with the SM noise -- (or indeed both).
\begin{table}
\begin{center}
\begin{tabular}{|c|c|c|c|c|c|c|c|}
\hline
$M_{Z'}\,[\text{GeV}]$&$g_{B-L}$&$g_{L(u\bar{u}Z')}$&$g_{R(u\bar{u}Z')}$&$g_{L(d\bar{d}Z')}$&$g_{R(d\bar{d}Z')}$&$g_{L(e\bar{e}Z')}$&$g_{R(e\bar{e}Z')}$\\
\hline
 3059.86&0.5&0.0750168& 0.049799& 0.0748826& 0.1001& 0.224916& 0.199698\\
\hline
2447.89&0.4&0.0583854& 0.0331175& 0.0581732& 0.0834411& 0.174944& 0.149676 \\
\hline
2019.51&0.33&0.0467655& 0.0214310& 0.0464498& 0.0717843& 0.139981& 0.114646 \\
\hline
1468.73&0.24&0.0319006& 0.0063741& 0.0312861& 0.0568126& 0.095087& 0.069561 \\
\hline
\end{tabular}
\caption{The four benchmark points for the narrow $Z'$ case with $\Gamma_{Z^\prime} \simeq 100$~GeV, $M_{\tilde{\nu}_{R_1}} \simeq M_{\tilde{\nu}_{R_2}}\simeq M_{\tilde{\nu}_{R_3}} \simeq 580$~GeV, $M_{\tilde{\nu}_{R_4}} \simeq M_{\tilde{\nu}_{R_5}}\simeq M_{\tilde{\nu}_{R_6}} \simeq 740$~GeV, $m_{\tilde{\chi}^\pm_{1,2}}\simeq 4, 0.9$~TeV, $m_{\tilde{\chi}^0_{1}}\simeq 440$~GeV and slepton masses of order $700$~GeV.}\label{table:narrow}
\end{center}
\end{table}
The profile of the cross section distributions are shown in Fig.~\ref{fig:Narrow_XS} for each of the four benchmarks. Clearly, we are here in the context of narrow resonances as $\Gamma_{Z^\prime} / M_{Z^\prime} \simeq 3 - 7 \%$.
Experimentally, narrow resonance searches are performed through a scan over the invariant mass distribution of the dilepton final-state system using the highest resolution possible. For our simulation, since a small number of event is expected in each bin, we have chosen to compute the significance of the signal produced by such a resonance using a Poisson statistics approach \cite{Curtis,Frodesen}.
The significances of the signal produced by the four benchmarks are shown in Fig.~\ref{fig:Narrow_Sig}:
even on the peak the significance of the signal would be below 1.

To continue, 
as we said that the BLSSM  provides a natural framework to explore the phenomenology of broad resonances through 
the opening of $Z^\prime$ decay channels into the SUSY sector, we also present benchmarks 
where the $Z'$ resonance is very broad.
On the experimental side, the wide resonance case is studied through a 'counting strategy' approach, that is, by looking for an excess in the number of events starting from a certain 
mass threshold (typically just above a control region) up to the end of the invariant mass spectrum.

\begin{table}
\begin{center}
\begin{tabular}{|c|c|c|c|c|c|c|c|}
\hline
$M_{Z'}\,[\text{GeV}]$&$g_{B-L}$&$g_{L(u\bar{u}Z')}$&$g_{R(u\bar{u}Z')}$&$g_{L(d\bar{d}Z')}$&$g_{R(d\bar{d}Z')}$&$g_{L(e\bar{e}Z')}$&$g_{R(e\bar{e}Z')}$\\
\hline
3041.53& 0.5& 0.0215611& 0.167177& 0.0198384& 0.208576& 0.0629607& 0.125777\\
\hline
2433.22& 0.4& 0.00534572& 0.184038& 0.00261583& 0.192& 0.0133073& 0.176077\\
\hline
2008.43& 0.33& 0.00572072& 0.195963& 0.00978994& 0.180452& 0.0212314& 0.211474 \\
\hline
1520.76& 0.25& 0.017618& 0.209908& 0.0248896& 0.167401& 0.0601255& 0.252416 \\
\hline
\end{tabular}
\caption{The four benchmark points for the wide $Z'$ case with $\Gamma_{Z^\prime} \simeq 810$~GeV, $M_{\tilde{\nu}_{R_1}} \simeq M_{\tilde{\nu}_{R_2}}\simeq M_{\tilde{\nu}_{R_3}} \simeq 610$~GeV, $M_{\tilde{\nu}_{R_4}} \simeq M_{\tilde{\nu}_{R_5}}\simeq M_{\tilde{\nu}_{R_6}} \simeq 760$~GeV, $m_{\tilde{\chi}^\pm_{1,2}}\simeq 4, 0.9$~TeV, $m_{\tilde{\chi}^0_{1}}\simeq 340$~GeV and slepton masses of order $700$~GeV.}\label{table:wide}
\end{center}
\end{table}

\begin{figure}[t]
\begin{center}
\includegraphics[width=0.8\textwidth]{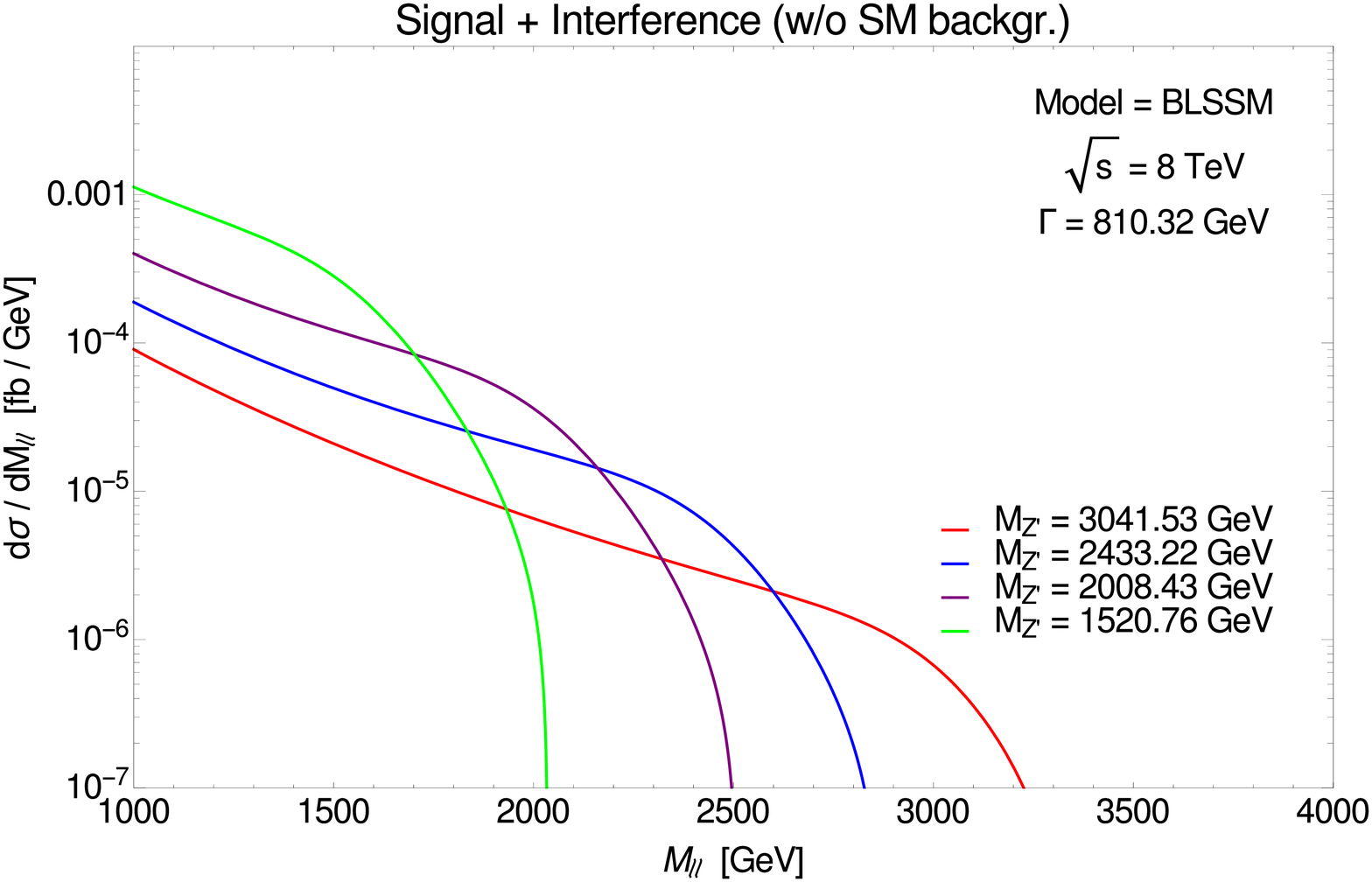}
\caption{Differential cross section distributions at LO in DY for the four benchmarks in Tab.~\ref{table:wide}.}
\label{fig:Wide_XS}
\end{center}
\end{figure}

\begin{figure}[t]
\begin{center}
\includegraphics[width=0.8\textwidth]{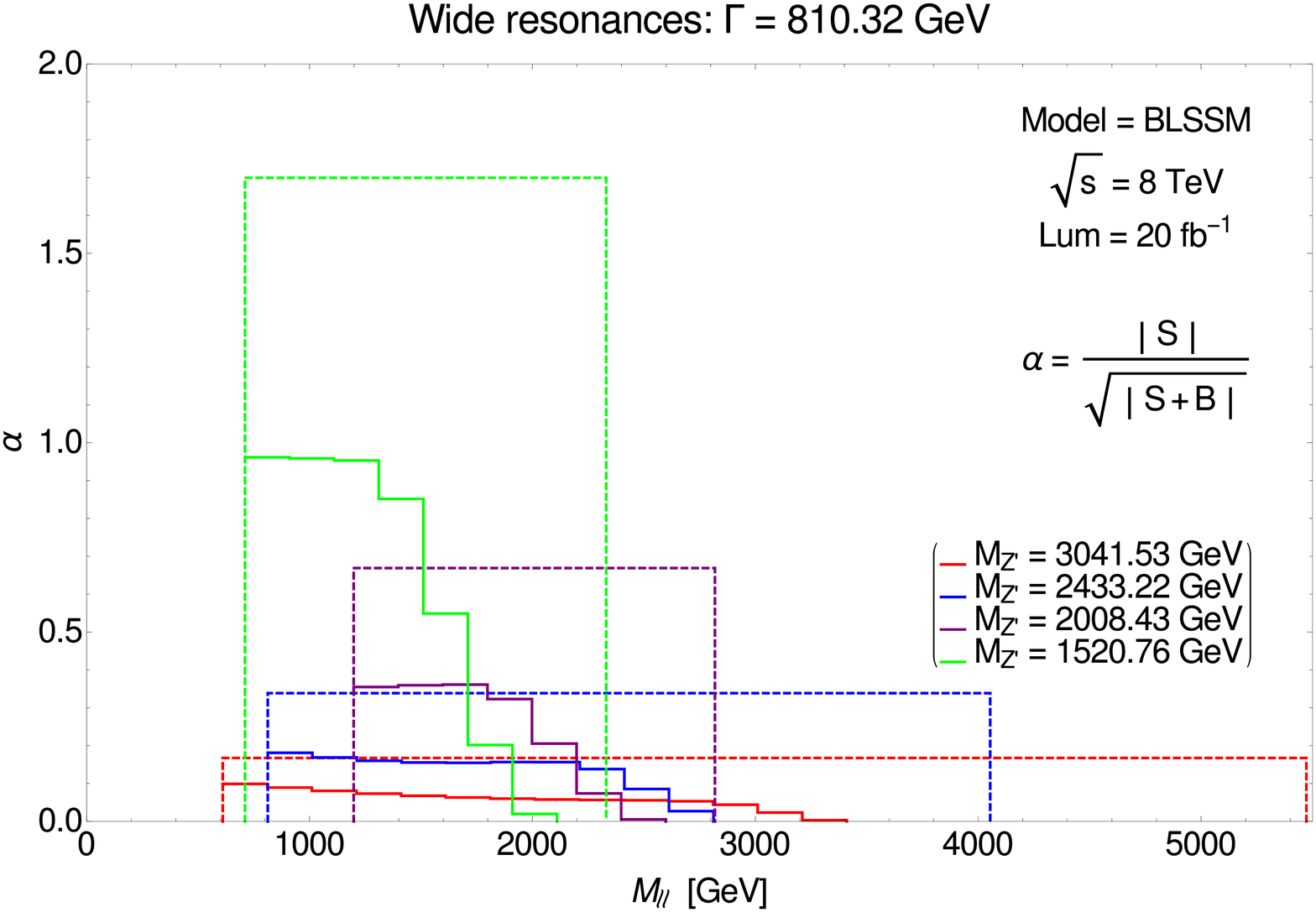}
\caption{Significance ($\alpha$) of the $Z^\prime$ signal in the dilepton channel for the four benchmarks in Tab.~\ref{table:wide}. We are also presenting the integrated significance in an extended invariant mass region (large bin integration). Gaussian statistics has been assumed.}
\label{fig:Wide_Sig}
\end{center}
\end{figure}

The four benchmarks that we have identified to study the wide $Z^\prime$ case (see Tab.~\ref{table:wide}) feature a ratio $\Gamma_{Z'} / M_{Z'} > 25 \%$ and their cross section distributions are seen  in Fig.~\ref{fig:Wide_XS}. Following the experimental procedure adopted in this context, the excesses of events we find for these benchmarks are not significant 
either (see Fig.~\ref{fig:Wide_Sig}). Here, Gauss statistics has been adopted. 

Hence, the two groups of benchmarks that we have introduced will constitute the framework of the analysis we are presenting in this paper and, at the same time, they represent
realistic scenarios not yet ruled out by up-to-date experimental constraints. In the definition of such BLSSM benchmarks we have used SARAH \cite{florianSARAH} and SPheno \cite{PorodSPheno,florianSPheno} to build the BLSSM and calculate masses, couplings and BRs. Then, the matrix-element calculation and parton level $S$  and $B$ events were derived from \textsf{MadGraph5} \cite{Madgraph5} whereas, for showering and hadronization, we have used \textsf{PYTHIA}  \cite{Sjostrand:2006za}. Further,  we have performed a fast detector simulation with \textsf{PGS4} \cite{PGS4}. Finally, we have manipulated the Monte Carlo (MC) data with \textsf{MadAnalysis5} \cite{Madanalysis5}. 

\section{Mono-jet signal}
\label{sec:monojet}

Using two points with $M_{Z'}$ in the $\sim 2.4$~TeV (narrow $Z'$ case) and $\sim2$~TeV (wide $Z'$ case) range
which were shown in Sec.~\ref{sec:limits}, we study the detection of the aforementioned BLSSM invisible final states in mono-jet searches at the LHC. As intimate, this signature  is often dominated by sneutrino decays, alongside the 
more traditional neutrino and neutralino modes, as follows (see Fig.~\ref{fig:mono_feynman})

\begin{eqnarray}
pp \to Z'(\to\tilde\nu_R\tilde\nu^*_R\to \tilde\chi^{0}_{1}\tilde\chi^{0}_{1}\,\nu\bar\nu)+j,~~ Z'(\to\nu\bar{\nu})+j,~~ Z'(\to\tilde{\chi}^{0}_{1}\tilde{\chi}^{0}_{1})+j,
\label{monojetsignal}
\end{eqnarray}
where $j$ represents a jet and the $\nu$ and $\tilde{\chi}^0_1$ states are invisible to the detector, thereby producing missing transverse energy, $\slashed E_T$, in it.

The SM backgrounds with respect to these mono-jet processes are dominated by the following channels: (i) the irreducible background $pp \to Z(\to \nu\bar{\nu})+j$, which is the main one because it has the same topology as our signals; (ii) $pp \to W(\to \ell \nu)+j$ ($\ell=e,\mu,\tau$), this process fakes the signal only when the charged lepton is outside the acceptance of the detector or close to the jet; (iii) $pp \to W(\to \tau \nu)+j$, this process may fake the signal since a secondary jet from hadronic tau decays tend to localize on the side opposite to $\slashed E_T$; (iv) $pp \to t\bar{t}$, this process may resemble the signal but also contains extra jets and leptons, which  allow one to highly suppress it by applying $b$-jet and lepton vetoes; (v) the di-boson background $pp \to ZZ(\to 2\nu 2\bar{\nu}) +j$,  which is generically suppressed due to its small cross section at production level but topologically mimic our signals rather well.

\begin{figure}[t]
\includegraphics[width=8cm,height=7cm]{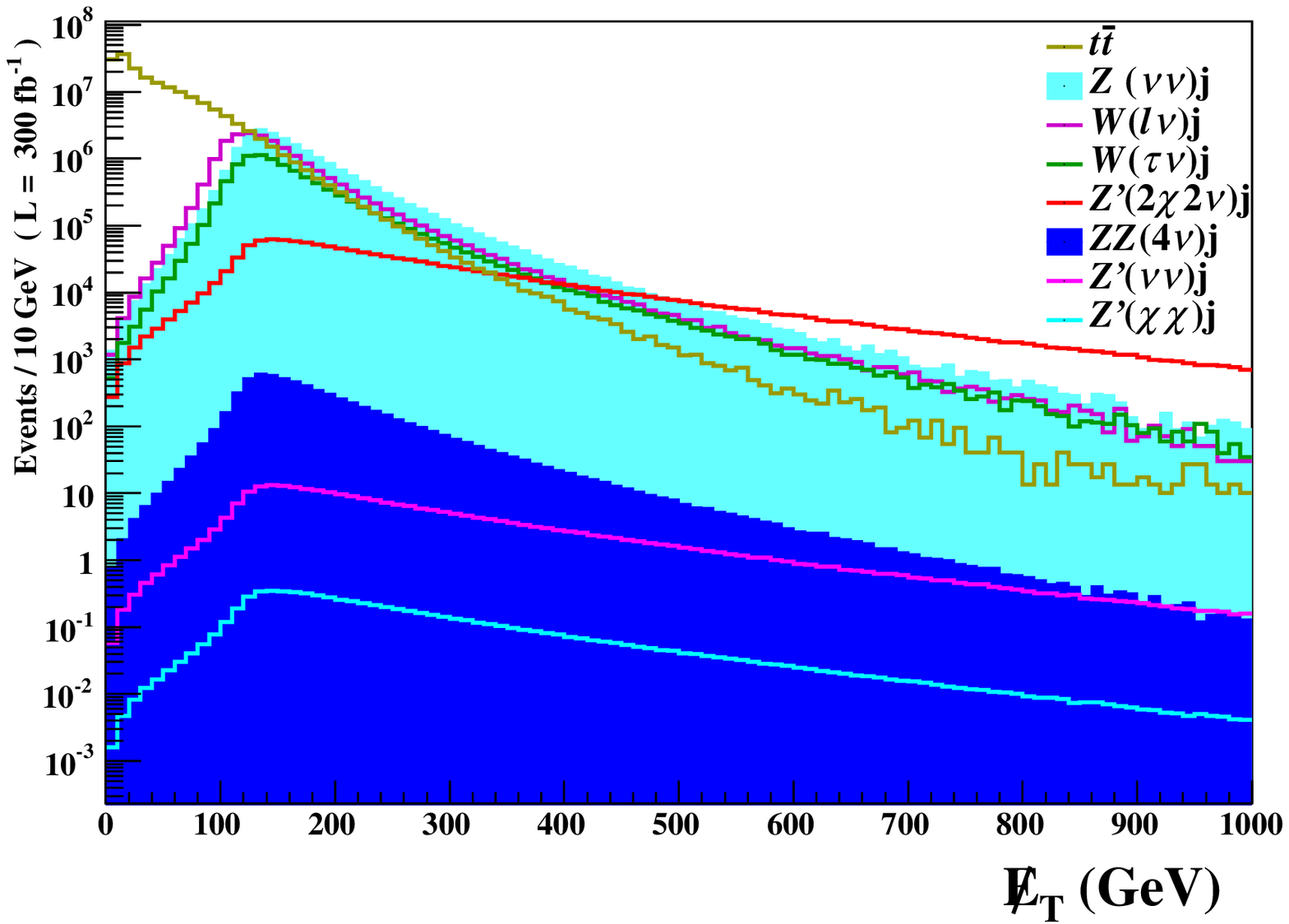} 
\includegraphics[width=8cm,height=7cm]{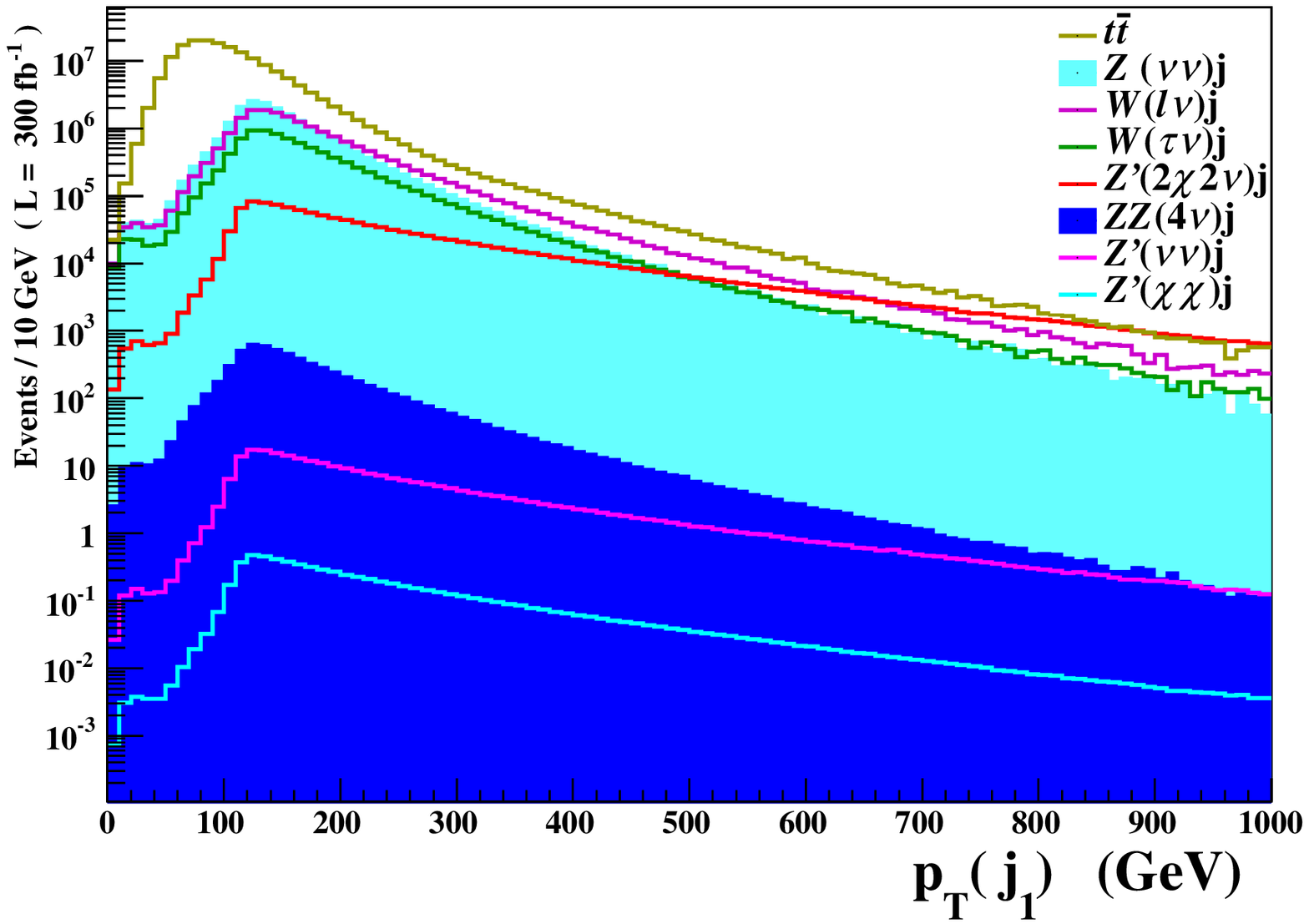}
\caption{(Left panel) Number of events versus the missing transverse energy. (Right panel) Number of events versus the transverse  momentum of the leading jet.
Both plots are presented before selection (i.e., detector level) cuts but after the parton level (i.e., MC generation) cuts $\slashed E_T>100$~GeV and $p_T(j_1)>120$~GeV.
Rates are given at 14 TeV for an integrated luminosity of 300~fb$^{-1}$. Here, $M_{Z'}\simeq 2448$~GeV and $g_{B-L}=0.4$ (narrow $Z'$ case).}
\label{fig1}
\end{figure}

\begin{figure}[t]
\includegraphics[width=8cm,height=7cm]{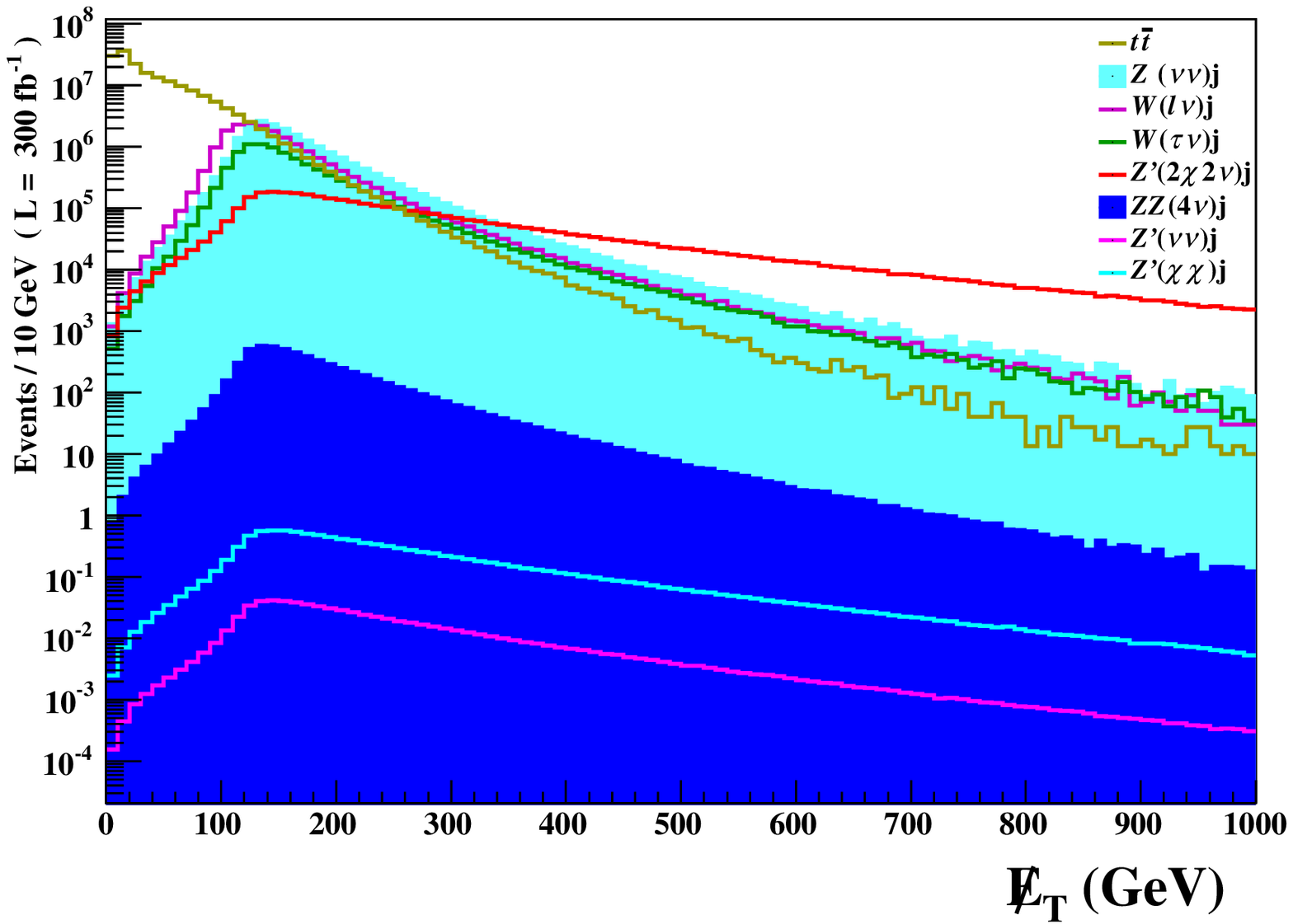} 
\includegraphics[width=8cm,height=7cm]{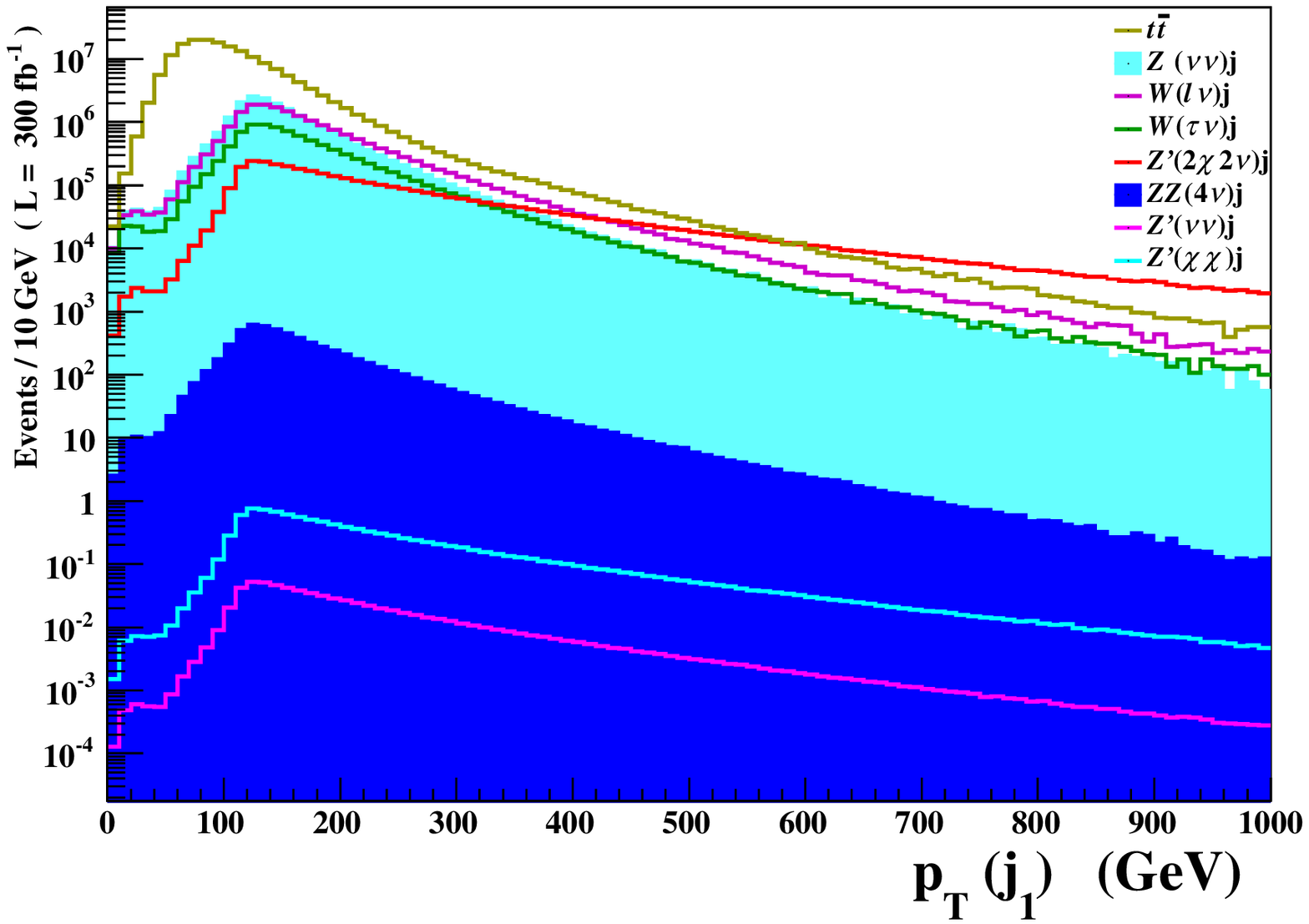}
\caption{(Left panel) Number of events versus the missing transverse energy. (Right panel) Number of events versus the transverse  momentum of the leading jet.
Both plots are presented before selection (i.e., detector level) cuts but after the parton level (i.e., MC generation) cuts $\slashed E_T>100$~GeV and $p_T(j_1)>120$~GeV.
Rates are given at 14 TeV for an integrated luminosity of 300~fb$^{-1}$. Here, $M_{Z'}\simeq 2008$~GeV and $g_{B-L}=0.33$ (wide $Z'$ case).}
\label{fig2}
\end{figure}

For MC efficiency and in order to obtain reasonable statistics, we have applied a parton level (generation) cut of $p_T(j_1) > 120$~GeV (on the highest transverse momentum jet $j_1$) for all signals and backgrounds \cite{Baer:2014cua,Han:2013usa}. According to the estimation of the QCD background based on the full detector simulation of Refs.~\cite{Aad:2009wy,allanach}, in the SUSY mono-jets analysis at 14~TeV LHC, the multi-jet background can be reduced to a negligible level by requiring a large $\slashed E_T$ cut, so we have applied another parton level (generation) cut of $\slashed E_T > 100$~GeV for both signals and backgrounds. As explained in Sec.~\ref{sec:intro}, owing to the large mass of the intervening $Z'$, we can in the end afford a rather stiff cut in $\slashed E_T$ in order to enhance $S/\sqrt{B}$. This is done by setting  $\slashed E_T > 500$~GeV \cite{Drees:2012dd}. This choice is justified by
Figs.~\ref{fig1}--\ref{fig2}. Here, for the case of a narrow and wide $Z'$, respectively, we show the $p_T(j_1)$ and $\slashed E_T$ distributions of  all signals and backgrounds. From the left panel one observes that the signals have indeed a much larger $\slashed E_T$ than the backgrounds. Thus, a hard cut on $\slashed E_T$ will be effective in order to enhance 
the $S/\sqrt B$ ratio. We supplement this by also requiring $p_T(j_1)>500$~GeV as the signals have harder $p_T(j_1)$ spectra than the backgrounds, unsurprisingly.


In Tabs.~\ref{tab:monojet1}--\ref{tab:monojet2}, again for a narrow and wide $Z'$, respectively, the resulting cut flow for signal and background events is presented at 14~TeV with an integrated luminosity of 300~fb$^{-1}$. After the cuts $p_T(j_1)>500$ GeV and $\slashed E_T >500$ GeV, all the backgrounds are reduced under the dominated sneutrino signal. In particular, notice the effectiveness of the lepton and and $b$-jet vetoes, which have suppressed $t\bar{t}$ (and $Wj$ as well) by more than two orders of magnitude \cite{lhc}.

\begin{table}[t]
{\small\fontsize{7}{7}\selectfont{
\begin{tabular}{|@{\hspace{0.02cm}}c@{\hspace{0.02cm}}||@{\hspace{0.02cm}}c@{\hspace{0.02cm}}|@{\hspace{0.02cm}}c@{\hspace{0.02cm}}|@{\hspace{0.02cm}}c@{\hspace{0.02cm}}|@{\hspace{0.02cm}}c@{\hspace{0.02cm}}|@{\hspace{0.02cm}}c@{\hspace{0.02cm}}|@{\hspace{0.02cm}}c@{\hspace{0.02cm}}|@{\hspace{0.02cm}}c@{\hspace{0.02cm}}|@{\hspace{0.02cm}}c@{\hspace{0.02cm}}|@{\hspace{0.02cm}}c@{\hspace{0.02cm}}|}
\hline
\multicolumn{2}{|@{\hspace{0.02cm}}c@{\hspace{0.02cm}}|}{}&\multicolumn{5}{c|}{Backgrounds}&\multicolumn{3}{c|}{Signals}\\
\hline
\multicolumn{2}{|@{\hspace{0.02cm}}c@{\hspace{0.02cm}}|}{Process}& $Z(\nu\bar{\nu})j$ & $W(l\nu_l)j$ & $W(\tau\nu_\tau)j$ &$t\bar{t}$& $ZZj$  
&$Z'(2\tilde\chi \,2\nu)j$&$Z'(\nu\bar{\nu})j$&$Z'(\tilde\chi\tilde\chi)j$ \\
\hline
\hline
\multicolumn{2}{|@{\hspace{0.02cm}}c@{\hspace{0.02cm}}|}{Before cuts} &21573000 & 19248000 & 9390000 & 179058000 & 6621 & 1334400 & 278 & 7.54  \\
\hline
\multirow{9}{*}{\rotatebox{90}{Cut}}&(1) &16823567 \!$\pm$\! 1924& 15817945 \!$\pm$\! 1678& 7719914 \!$\pm$\! 1171& 151390826 \!$\pm$\! 4836& 5732 \!$\pm$\! 28& 1219314 \!$\pm$\! 324& 255 \!$\pm$\! 4.68& 6.895 \!$\pm$\! 0.77 \\
\cline{2-10}
&(2)&65275 \!$\pm$\! 255& 135191 \!$\pm$\! 366& 65423 \!$\pm$\! 254& 298430 \!$\pm$\! 545& 73 \!$\pm$\! 8.5& 130636 \!$\pm$\! 343& 27 \!$\pm$\! 4.95& 0.741 \!$\pm$\! 0.82  \\
\cline{2-10}
&(3)&45530 \!$\pm$\! 213& 32569 \!$\pm$\! 180& 27102 \!$\pm$\! 164& 6836.8 \!$\pm$\! 82.7& 55.6 \!$\pm$\! 7.43& 118456 \!$\pm$\! 328& 25 \!$\pm$\! 4.74& 0.672 \!$\pm$\! 0.78  \\
\cline{2-10}
&(4)&14283 \!$\pm$\! 119& 10566 \!$\pm$\! 102& 8668.5 \!$\pm$\! 93.1& 2808 \!$\pm$\! 53& 16.5 \!$\pm$\! 4.06& 35424 \!$\pm$\! 185& 7.4 \!$\pm$\! 2.68& 0.201 \!$\pm$\! 0.44 \\
\cline{2-10}
&(5)&10831 \!$\pm$\! 104& 7395.3 \!$\pm$\! 86& 6088.7 \!$\pm$\! 78& 881.7 \!$\pm$\! 29.7& 12.2 \!$\pm$\! 3.49& 23330 \!$\pm$\! 151& 4.9 \!$\pm$\! 2.18& 0.132 \!$\pm$\! 0.36  \\
\cline{2-10}
&(6)&8992.5 \!$\pm$\! 94.8& 6007.4 \!$\pm$\! 77.5& 4699.9 \!$\pm$\! 68.5& 379.8 \!$\pm$\! 19.5& 9.79 \!$\pm$\! 3.13& 18806 \!$\pm$\! 136& 3.9 \!$\pm$\! 1.96& 0.107 \!$\pm$\! 0.33  \\
\cline{2-10}
&(7)&8969.8 \!$\pm$\! 94.7& 3343.1 \!$\pm$\! 57.8& 3929 \!$\pm$\! 62.7& 257.7 \!$\pm$\! 16.1& 9.78 \!$\pm$\! 3.12& 18786 \!$\pm$\! 136& 3.9 \!$\pm$\! 1.96& 0.107 \!$\pm$\! 0.32  \\
\cline{2-10}
&(8)&8969.8 \!$\pm$\! 94.7& 871.2 \!$\pm$\! 29.5& 3207.4 \!$\pm$\! 56.6& 176.3 \!$\pm$\! 13.3& 9.77 \!$\pm$\! 3.12& 18782 \!$\pm$\! 136& 3.9 \!$\pm$\! 1.96& 0.107 \!$\pm$\! 0.32  \\
\cline{2-10}
&(9)&8458.9 \!$\pm$\! 92& 790.2 \!$\pm$\! 28.1& 1378.8 \!$\pm$\! 37.1& 81.39 \!$\pm$\! 9.02& 9.21 \!$\pm$\! 3.03& 17878 \!$\pm$\! 132& 3.7 \!$\pm$\! 1.92& 0.102 \!$\pm$\! 0.32  \\
\cline{2-10}
&(10)&8152.3 \!$\pm$\! 90.3& 769.9 \!$\pm$\! 27.7& 1334.4 \!$\pm$\! 36.5& 54.26 \!$\pm$\! 7.37& 8.8 \!$\pm$\! 2.96& 17357 \!$\pm$\! 130& 3.6 \!$\pm$\! 1.89& 0.098 \!$\pm$\! 0.31  \\
\hline
\end{tabular}}}
\caption{The cut flow on signal and background events after requiring the parton level cuts $\slashed E_T>100$~GeV and $p_T(j_1)>120$~GeV for $M_{Z'}\simeq 2448$~GeV and $g_{B-L}=0.4$ (narrow $Z'$ case) in the mono-jet channel at $\sqrt s=14$~TeV with 
${\cal L}dt= 300$~fb$^{-1}$: (1) $n(\text{jets})\geq 1$ with $|\eta(j_1)|<2$; (2) $p_T(j_1)> 500$~GeV; (3) $\met > 500$~GeV; (4) $\Delta \phi(j_2,\met) > 0.5$; (5) veto on $p_T(j_2)> 100$~GeV, $|\eta(j_2)|< 2$; (6) veto on $p_T(j_3)> 30$~GeV, $|\eta(j_3)|< 4.5$; (7) veto on $e$; (8) veto on $\mu$; (9) veto on $\tau$-jets; (10) veto on $b$-jets.}
\label{tab:monojet1} 
\end{table}
\begin{table}[t]
{\small\fontsize{7}{7}\selectfont{
\begin{tabular}{|@{\hspace{0.02cm}}c@{\hspace{0.02cm}}||@{\hspace{0.02cm}}c@{\hspace{0.02cm}}|@{\hspace{0.02cm}}c@{\hspace{0.02cm}}|@{\hspace{0.02cm}}c@{\hspace{0.02cm}}|@{\hspace{0.02cm}}c@{\hspace{0.02cm}}|@{\hspace{0.02cm}}c@{\hspace{0.02cm}}|@{\hspace{0.02cm}}c@{\hspace{0.02cm}}|@{\hspace{0.02cm}}c@{\hspace{0.02cm}}|@{\hspace{0.02cm}}c@{\hspace{0.02cm}}|@{\hspace{0.02cm}}c@{\hspace{0.02cm}}|}
\hline
\multicolumn{2}{|@{\hspace{0.02cm}}c@{\hspace{0.02cm}}|}{}&\multicolumn{5}{c|}{Backgrounds}&\multicolumn{3}{c|}{Signals}\\
\hline
\multicolumn{2}{|@{\hspace{0.02cm}}c@{\hspace{0.02cm}}|}{ Process}& $Z(\nu\bar{\nu})j$ & $W(l\nu_l)j$ & $W(\tau\nu_\tau)j$ &$t\bar{t}$& $ZZj$  
&$Z'(2\tilde\chi \,2\nu)j$&$Z'(\nu\bar{\nu})j$&$Z'(\tilde\chi\tilde\chi)j$ \\
\hline
\hline
\multicolumn{2}{|@{\hspace{0.02cm}}c@{\hspace{0.02cm}}|}{Before cuts} &21573000 & 19248000 & 9390000 & 179058000 & 6621 & 3976451 & 0.788 & 11.8  \\
\hline
\multirow{9}{*}{\rotatebox{90}{Cut}}&(1) &16823567 \!$\pm$\! 1924& 15817945 \!$\pm$\! 1678& 7719914 \!$\pm$\! 1171& 151390826 \!$\pm$\! 4836& 5732 \!$\pm$\! 28& 3572791 \!$\pm$\! 602& 0.71 \!$\pm$\! 0.27& 10.7 \!$\pm$\! 1.01 \\
\cline{2-10}
&(2)&65275 \!$\pm$\! 255& 135191 \!$\pm$\! 366& 65423 \!$\pm$\! 254& 298430 \!$\pm$\! 545& 73 \!$\pm$\! 8.5& 393706 \!$\pm$\! 595& 0.06 \!$\pm$\! 0.24& 1.06 \!$\pm$\! 0.98  \\
\cline{2-10}
&(3)&45530 \!$\pm$\! 213& 32569 \!$\pm$\! 180& 27102 \!$\pm$\! 164& 6836.8 \!$\pm$\! 82.7& 55.6 \!$\pm$\! 7.43& 353849 \!$\pm$\! 567& 0.05 \!$\pm$\! 0.23& 0.94 \!$\pm$\! 0.93  \\
\cline{2-10}
&(4)&14283 \!$\pm$\! 119& 10566 \!$\pm$\! 102& 8668.5 \!$\pm$\! 93.1& 2808 \!$\pm$\! 53& 16.5 \!$\pm$\! 4.06& 107606 \!$\pm$\! 323& 0.02 \!$\pm$\! 0.13& 0.29 \!$\pm$\! 0.53 \\
\cline{2-10}
&(5)&10831 \!$\pm$\! 104& 7395.3 \!$\pm$\! 86& 6088.7 \!$\pm$\! 78& 881.7 \!$\pm$\! 29.7& 12.2 \!$\pm$\! 3.49& 70542 \!$\pm$\! 263& 0.01 \!$\pm$\! 0.10& 0.19 \!$\pm$\! 0.43  \\
\cline{2-10}
&(6)&8992.5 \!$\pm$\! 94.8& 6007.4 \!$\pm$\! 77.5& 4699.9 \!$\pm$\! 68.5& 379.8 \!$\pm$\! 19.5& 9.79 \!$\pm$\! 3.13& 56097 \!$\pm$\! 235& 0.01 \!$\pm$\! 0.09& 0.15 \!$\pm$\! 0.39  \\
\cline{2-10}
&(7)&8969.8 \!$\pm$\! 94.7& 3343.1 \!$\pm$\! 57.8& 3929 \!$\pm$\! 62.7& 257.7 \!$\pm$\! 16.1& 9.78 \!$\pm$\! 3.12& 56030 \!$\pm$\! 235& 0.01 \!$\pm$\! 0.09& 0.15 \!$\pm$\! 0.39  \\
\cline{2-10}
&(8)&8969.8 \!$\pm$\! 94.7& 871.2 \!$\pm$\! 29.5& 3207.4 \!$\pm$\! 56.6& 176.3 \!$\pm$\! 13.3& 9.77 \!$\pm$\! 3.12& 56013 \!$\pm$\! 234& 0.01 \!$\pm$\! 0.09& 0.15 \!$\pm$\! 0.39  \\
\cline{2-10}
&(9)&8458.9 \!$\pm$\! 92& 790.2 \!$\pm$\! 28.1& 1378.8 \!$\pm$\! 37.1& 81.39 \!$\pm$\! 9.02& 9.21 \!$\pm$\! 3.03& 53221 \!$\pm$\! 229& 0.01 \!$\pm$\! 0.09& 0.14 \!$\pm$\! 0.38  \\
\cline{2-10}
&(10)&8152.3 \!$\pm$\! 90.3& 769.9 \!$\pm$\! 27.7& 1334.4 \!$\pm$\! 36.5& 54.26 \!$\pm$\! 7.37& 8.8 \!$\pm$\! 2.96& 51455 \!$\pm$\! 225& 0.01 \!$\pm$\! 0.09& 0.14 \!$\pm$\! 0.37  \\
\hline
\end{tabular}}}
\caption{The cut flow on signal and background events after requiring the parton level cuts $\slashed E_T>100$~GeV and $p_T(j_1)>120$~GeV for $M_{Z'}\simeq 2008$~GeV and $g_{B-L}=0.33$ (wide $Z'$ case) in the mono-jet channel at $\sqrt s=14$~TeV with 
${\cal L}dt= 300$~fb$^{-1}$: (1) $n(\text{jets})\geq 1$ with $|\eta(j_1)|<2$; (2) $p_T(j_1)> 500$~GeV; (3) $\met > 500$~GeV; (4) $\Delta \phi(j_2,\met) > 0.5$; (5) veto on $p_T(j_2)> 100$~GeV, $|\eta(j_2)|< 2$; (6) veto on $p_T(j_3)> 30$~GeV, $|\eta(j_3)|< 4.5$; (7) veto on $e$; (8) veto on $\mu$; (9) veto on $\tau$-jets; (10) veto on $b$-jets.}
\label{tab:monojet2} 
\end{table}

\begin{figure}[t]
\begin{center}
\includegraphics[width=7.0cm,height=5.5cm]{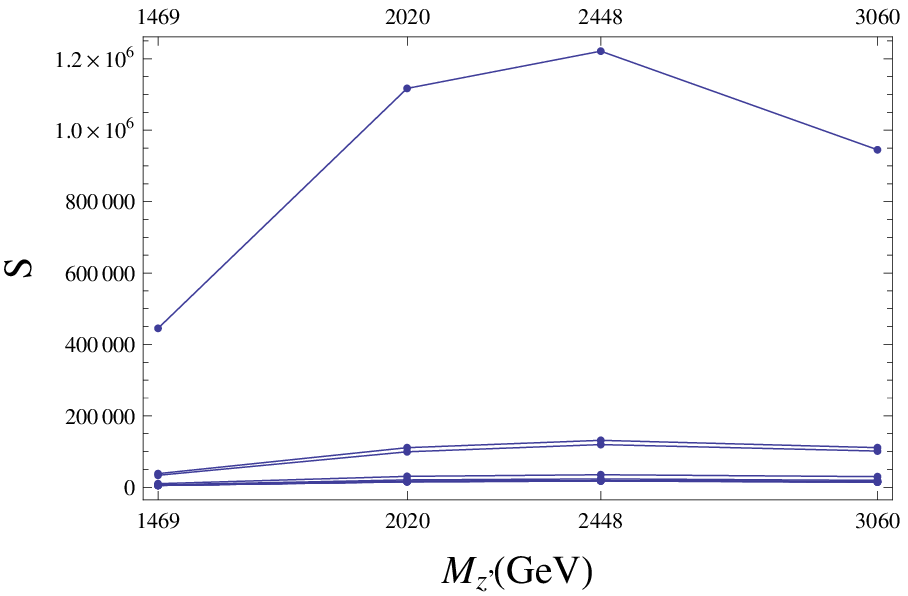}~\includegraphics[width=7.0cm,height=5.5cm]{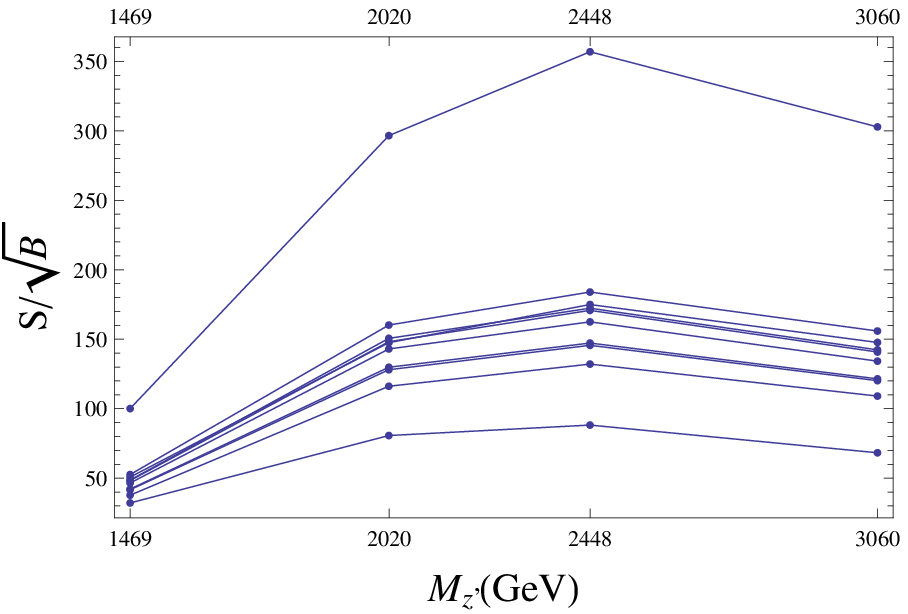}\\
\includegraphics[width=7.0cm,height=5.5cm]{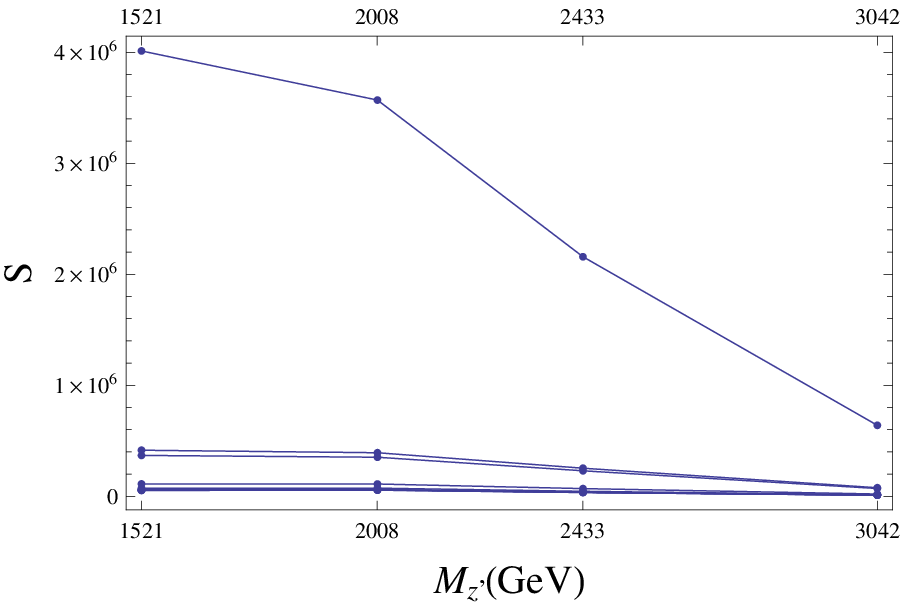}~\includegraphics[width=7.0cm,height=5.5cm]{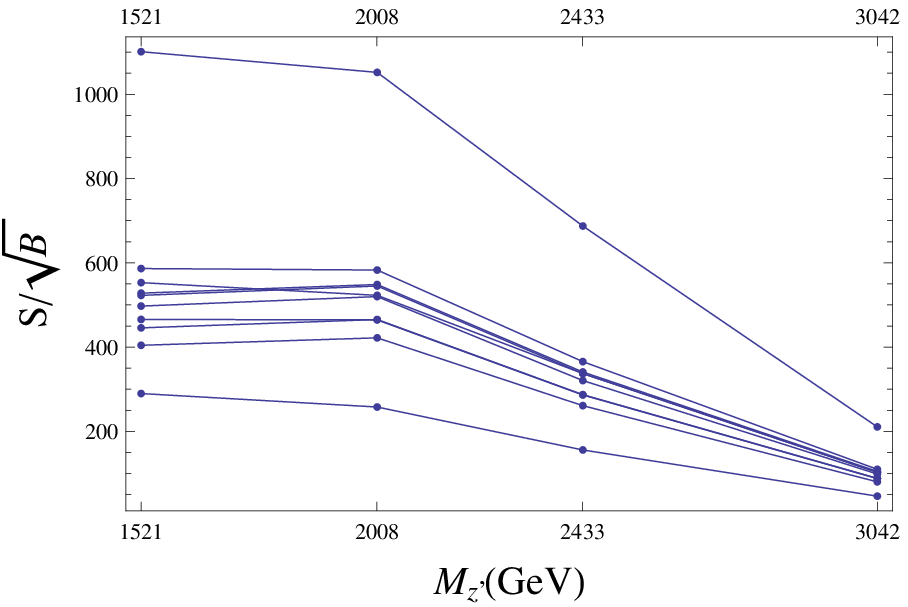}
\caption{(Top panel) Number of events from the sum of all signals $(S)$ versus $M_{Z'}$ and number of events from the sum of all signals divided by the square root of the total background $(S/\sqrt{B})$ versus $M_{Z'}$ for mono-jet in the narrow $Z'$ case. (Bottom panel) Number of events from the sum of all signals $(S)$ versus $M_{Z'}$ and number of events from the sum of all signals divided by the square root of the total background $(S/\sqrt{B})$ versus $M_{Z'}$ for mono-jet in the wide $Z'$ case. Rates are given at 14~TeV for an integrated luminosity of 300~fb$^{-1}$. } \label{monojetSBN}
\end{center}
\end{figure}

Finally, in Fig.~\ref{monojetSBN}, we present the signal rates and  significances, always at 300~fb$^{-1}$, for the
four benchmarks in Tabs.~\ref{table:narrow}--\ref{table:wide}, wherein any point corresponds to one of the
ten cuts in Tabs.~\ref{tab:monojet1}--\ref{tab:monojet2}. From such plots, it is evident that the mono-jet signal can be established for standard LHC conditions of energy and luminosity.

\section{Single-photon signal}
\label{sec:monophoton}

The Feynman diagrams contributing to the mono-photon process are similar to those we have seen in the mono-jet case
(modulus the absence of a sizable photon-induced contribution in the former, unlike the case of the 
substantial gluon-induced one in the latter), see Fig.~\ref{fig:mono_feynman} again.
We generate mono-photon events after requiring the following parton level (generation) cuts: $\met > 50$~GeV, $p_T(\gamma) > 40$~GeV and $p_T(j_1)> 25$~GeV (on the missing transverse energy, highest transverse momentum 
photon and jet, respectively). We also generate the background processes $Z(\to\nu\bar{\nu})\gamma$ and $W(\to \ell\nu_{\ell})\gamma$, where $\ell =e$,
$\mu$ or $\tau$ (as before).

\begin{figure}[t]
\includegraphics[width=8cm,height=7cm]{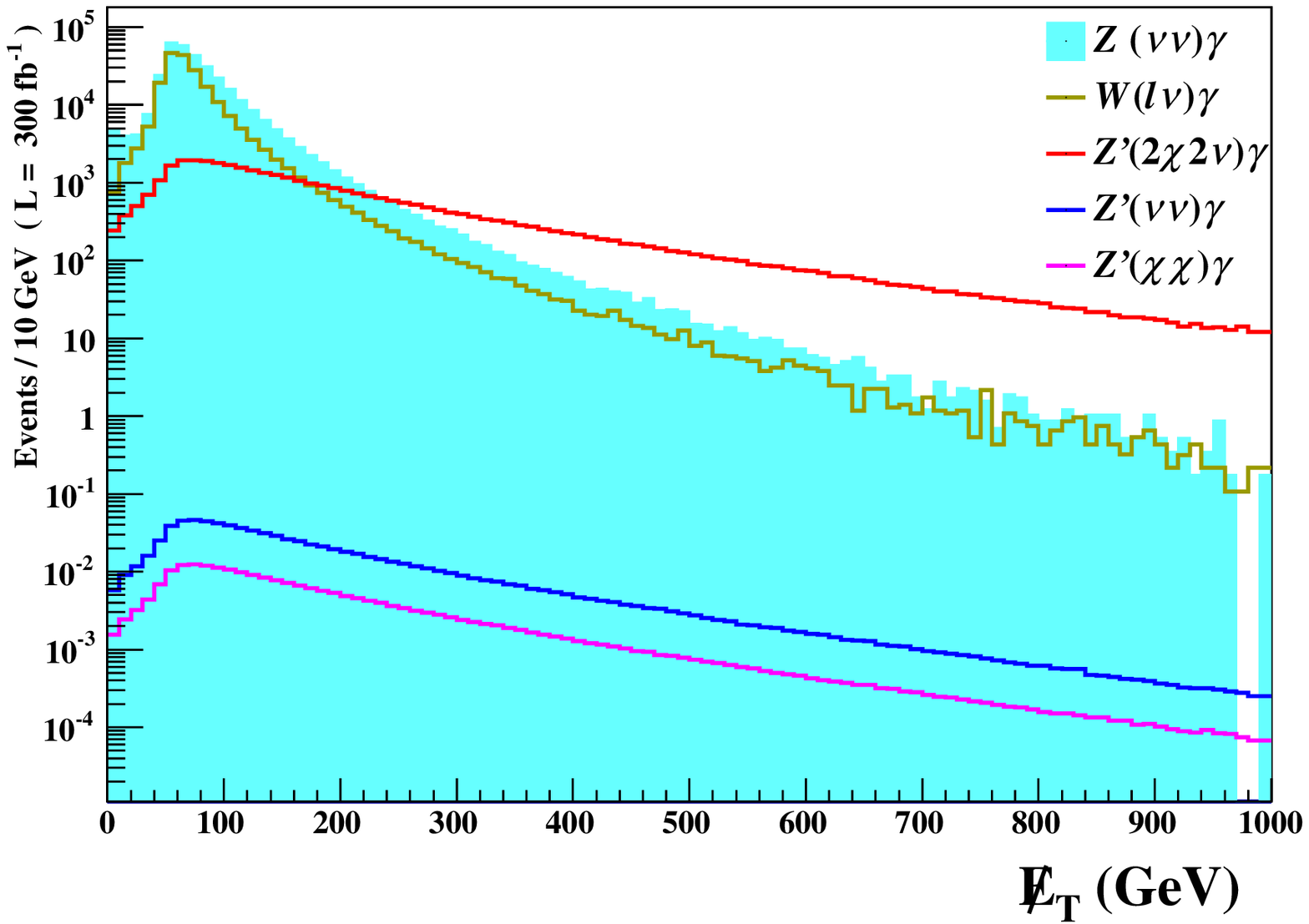} 
\includegraphics[width=8cm,height=7cm]{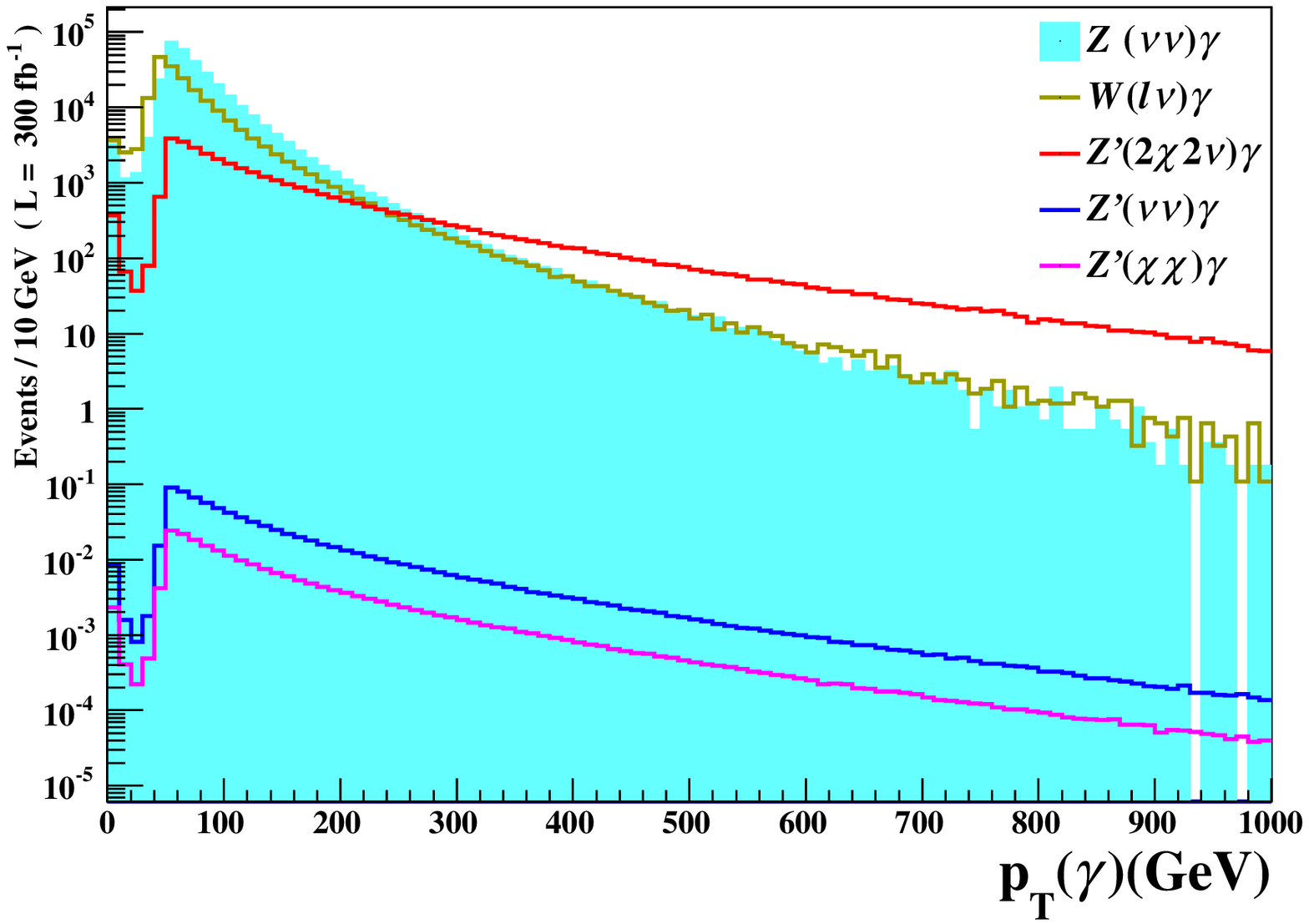}
\caption{(Left panel) Number of events versus the missing transverse energy. (Right panel) Number of events versus the transverse momentum of the photon.
Both plots are presented before selection (i.e., detector level) cuts but after the parton level (i.e., MC generation) cuts 
 $\slashed E_T>50$~GeV, $p_T(\gamma)>40$~GeV and $p_T(j_1)>30$~GeV.
Rates are given at 14 TeV for an integrated luminosity of 300~fb$^{-1}$. Here, $M_{Z'}\simeq 2448$~GeV and $g_{B-L}=0.4$ (narrow $Z'$ case).}
\label{fig3}
\end{figure}
\begin{figure}[t]
\includegraphics[width=8cm,height=7cm]{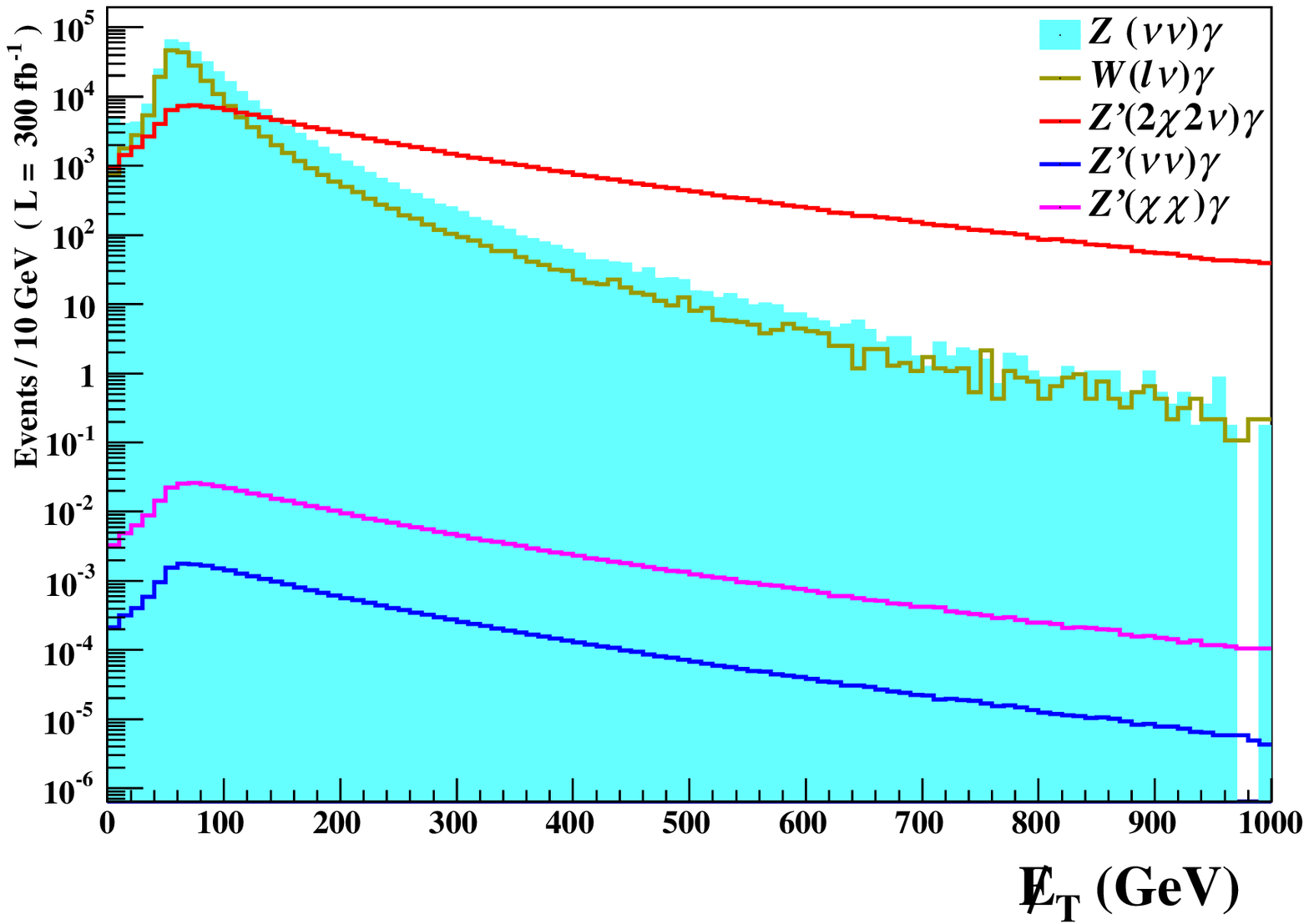} 
\includegraphics[width=8cm,height=7cm]{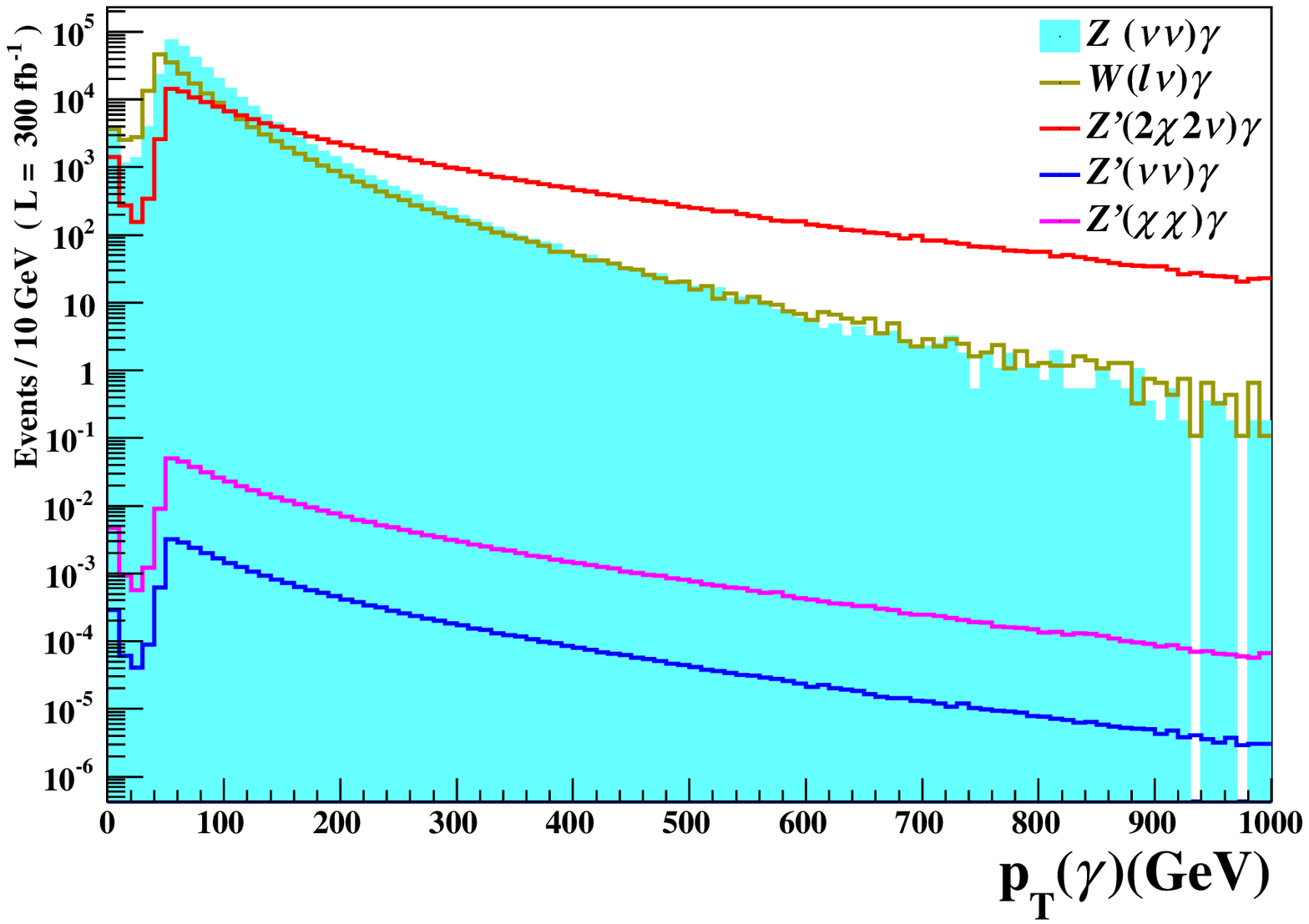}
\caption{(Left panel) Number of events versus the missing transverse energy. (Right panel) Number of events versus the transverse momentum of the photon. 
Both plots are presented before selection (i.e., detector level) cuts but after the parton level (i.e., MC generation) cuts 
 $\slashed E_T>50$~GeV, $p_T(\gamma)>40$~GeV and $p_T(j_1)>30$~GeV.
Rates are given at 14~TeV for an integrated luminosity of 300~fb$^{-1}$. Here, $M_{Z'}\simeq 2008$~GeV and $g_{B-L}=0.33$ (wide $Z'$ case).}
\label{fig4}
\end{figure}

\begin{figure}[t]
\begin{center}
\includegraphics[width=7.0cm,height=5.5cm]{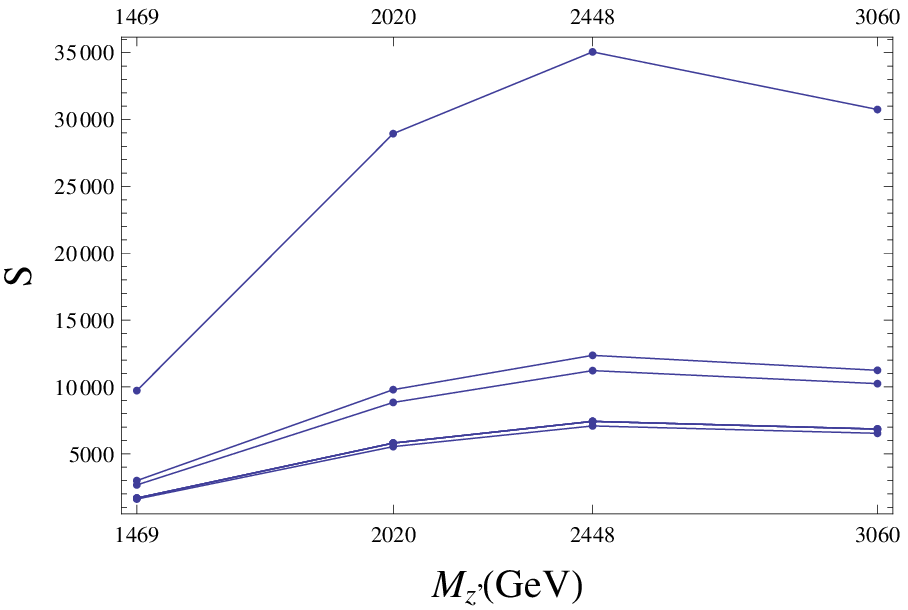}~\includegraphics[width=7.0cm,height=5.5cm]{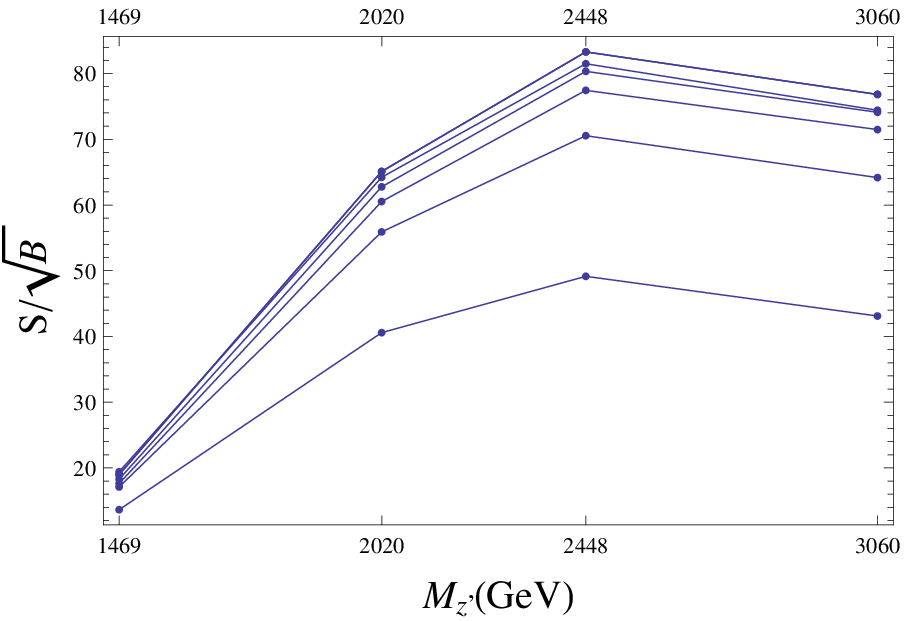}\\
\includegraphics[width=7.0cm,height=5.5cm]{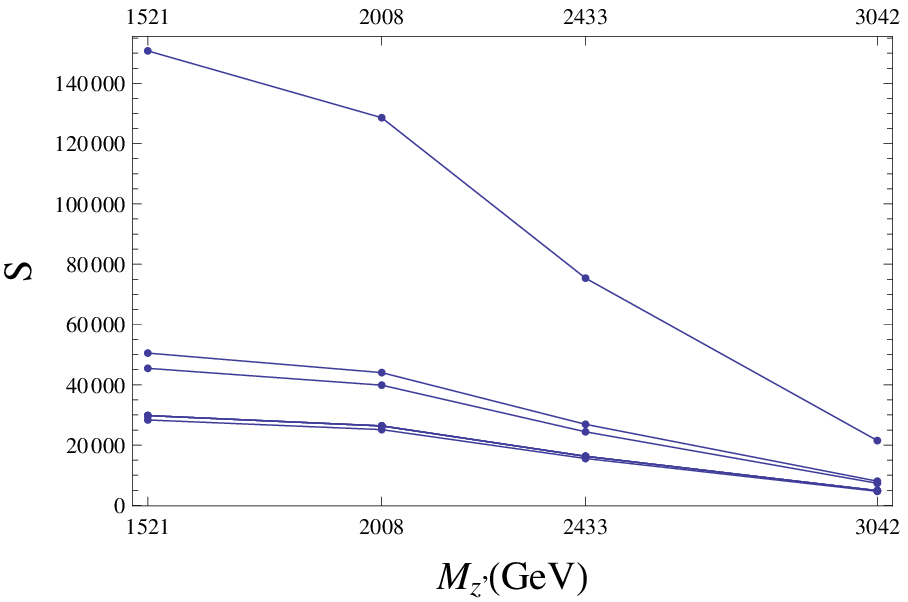}~\includegraphics[width=7.0cm,height=5.5cm]{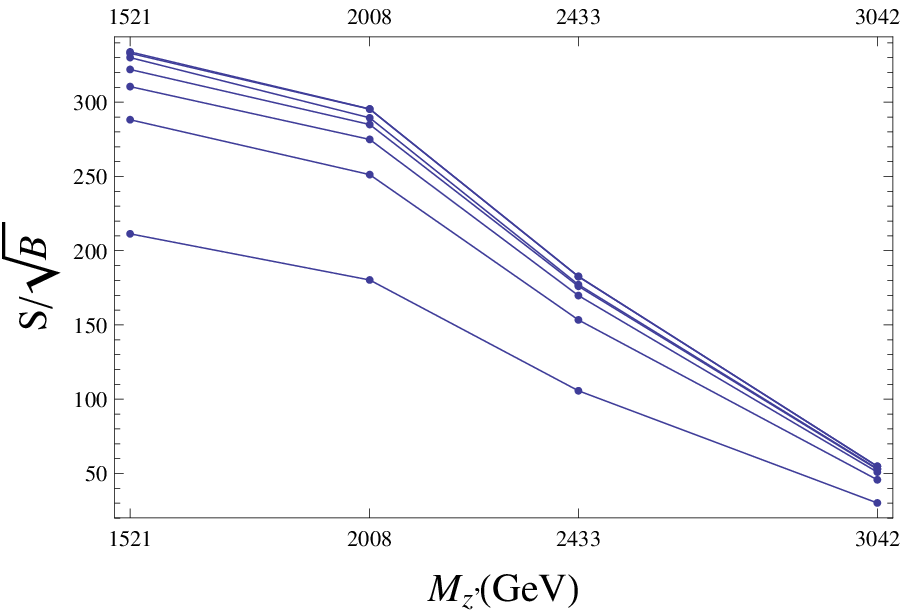}
\caption{(Top panel) Number of events from the sum of all signals $(S)$ versus $M_{Z'}$ and number of events from the sum of all signals divided by the square root of the total background $(S/\sqrt{B})$ versus $M_{Z'}$ for mono-photon in the narrow $Z'$ case. (Bottom panel) Number of events from the sum of all signals $(S)$ versus $M_{Z'}$ and number of events from the sum of all signals divided by the square root of the total background $(S/\sqrt{B})$ versus $M_{Z'}$ for mono-photon in the wide $Z'$ case. Rates are given at 14~TeV for an integrated luminosity of 300~fb$^{-1}$. } \label{monophotonSBN}
\end{center}
\end{figure}

In Figs.~\ref{fig3}--\ref{fig4}, the signal and background distributions in $p_T(\gamma )$ and $\met$
are shown for the two $Z'$ scenarios. We note from these figures that the shapes agree for large values of $p_T(\gamma)$ and $\met$, signalling a reduced jet activity in comparison to the mono-jet case. We also highlight that the
single-photon signal is somewhat stiffer than the mono-jet one in both dynamic variables with the backgrounds falling
more steeply in the former than in the latter case. This feature enables the single-photon signature to be somewhat 
relevant too for both discovery, albeit less so than the mono-jet one, with relative milder cuts in comparison, i.e.,
$\slashed E_T>150$~GeV and $p_T(\gamma)>150$~GeV \cite{Baer:2014cua,Aad:2012fw}. This is clear from Tabs.~\ref{tab:monophoton1}--\ref{tab:monophoton2} and Fig.~\ref{monophotonSBN}, where our cut flow is shown for our two customary
$Z'$ scenarios (narrow and wide) alongside the ensuing signal rates and significances. As intimated, prospects for detection are very positive.

\begin{table}
\begin{center}
\begin{tabular}{|@{\hspace{0.02cm}}c@{\hspace{0.02cm}}||@{\hspace{0.02cm}}c@{\hspace{0.02cm}}|@{\hspace{0.02cm}}c@{\hspace{0.02cm}}|@{\hspace{0.02cm}}c@{\hspace{0.02cm}}|@{\hspace{0.02cm}}c@{\hspace{0.02cm}}|@{\hspace{0.02cm}}c@{\hspace{0.02cm}}|c|}
\hline
\multicolumn{2}{|@{\hspace{0.02cm}}c@{\hspace{0.02cm}}|}{}&\multicolumn{2}{c|}{Backgrounds}&\multicolumn{3}{c|}{Signals}\\
\hline
\multicolumn{2}{|@{\hspace{0.02cm}}c@{\hspace{0.02cm}}|}{ Process}& $Z(\nu\bar{\nu})\gamma$ & $W(l\nu_l)\gamma$&$Z'(2\tilde\chi \,2\nu)\gamma$&$Z'(\nu\bar{\nu})\gamma$&$Z'(\tilde\chi\tilde\chi)\gamma$ \\
\hline
\hline
\multicolumn{2}{|@{\hspace{0.02cm}}c@{\hspace{0.02cm}}|}{Before cuts}& 332712 & 204644 & 37380 & 0.861 & 0.234  \\
\hline
\multirow{6}{*}{\rotatebox{90}{Cut}}&$n(\gamma)\geq 1$  & 316031 \!$\pm$\! 125& 192677 \!$\pm$\! 106& 34998 \!$\pm$\! 47.2& 0.806 \!$\pm$\! 0.227& 0.219 \!$\pm$\! 0.118 \\
\cline{2-7}
&$p_T(\gamma_1)> 150$~GeV  & 18576 \!$\pm$\! 132& 12146 \!$\pm$\! 106& 12357.8 \!$\pm$\! 91& 0.282 \!$\pm$\! 0.435& 0.0765 \!$\pm$\! 0.2268  \\
\cline{2-7}
&$\met > 150$~GeV  & 14681 \!$\pm$\! 118& 4287.3 \!$\pm$\! 64.8& 11202 \!$\pm$\! 88.6& 0.255 \!$\pm$\! 0.424& 0.0693 \!$\pm$\! 0.2208  \\
\cline{2-7}
&$n(j)\leq 1$, $|\eta(j)|< 4.5$  & 6819.7 \!$\pm$\! 81.7& 2388.2 \!$\pm$\! 48.6& 7415 \!$\pm$\! 77.1& 0.168 \!$\pm$\! 0.368& 0.0457 \!$\pm$\! 0.1917 \\
\cline{2-7}
&veto on $e$  & 6817.6 \!$\pm$\! 81.7& 1731.8 \!$\pm$\! 41.4& 7409.3 \!$\pm$\! 77.1& 0.168 \!$\pm$\! 0.368& 0.0456 \!$\pm$\! 0.1916  \\
\cline{2-7}
&veto on $\mu$  & 6817.6 \!$\pm$\! 81.7& 1132.5 \!$\pm$\! 33.6& 7407 \!$\pm$\! 77.1& 0.168 \!$\pm$\! 0.368& 0.0456 \!$\pm$\! 0.1916  \\
\cline{2-7}
&veto on $\tau$-jets  & 6479.8 \!$\pm$\! 79.7& 758 \!$\pm$\! 27.5& 7069 \!$\pm$\! 75.7& 0.161 \!$\pm$\! 0.631& 0.0435 \!$\pm$\! 0.1882 \\
\hline
\end{tabular}
\caption{The cut flow on signal and background events after requiring the parton level cuts $\slashed E_T>50$~GeV, $p_T(\gamma)>40$~GeV and $p_T(j_1)>30$~GeV for $M_{Z'}\simeq 2448$~GeV and $g_{B-L}=0.4$ (narrow $Z'$ case) in the single-photon channel at $\sqrt s=14$~TeV with ${\cal L}dt= 300$~fb$^{-1}$.}
\label{tab:monophoton1}
\end{center}
\end{table}

\begin{table}
\begin{center}
\begin{tabular}{|@{\hspace{0.02cm}}c@{\hspace{0.02cm}}||@{\hspace{0.02cm}}c@{\hspace{0.02cm}}|@{\hspace{0.02cm}}c@{\hspace{0.02cm}}|@{\hspace{0.02cm}}c@{\hspace{0.02cm}}|@{\hspace{0.02cm}}c@{\hspace{0.02cm}}|@{\hspace{0.02cm}}c@{\hspace{0.02cm}}|@{\hspace{0.02cm}}c@{\hspace{0.02cm}}|}
\hline
\multicolumn{2}{|@{\hspace{0.02cm}}c@{\hspace{0.02cm}}|}{}&\multicolumn{2}{c|}{Backgrounds}&\multicolumn{3}{c|}{Signals}\\
\hline
\multicolumn{2}{|@{\hspace{0.02cm}}c@{\hspace{0.02cm}}|}{ Process}& $Z(\nu\bar{\nu})\gamma$ & $W(l\nu_l)\gamma$&$Z'(2\tilde\chi \,2\nu)\gamma$&$Z'(\nu\bar{\nu})\gamma$&$Z'(\tilde\chi\tilde\chi)\gamma$ \\
\hline
\hline
\multicolumn{2}{|@{\hspace{0.02cm}}c@{\hspace{0.02cm}}|}{Before cuts}& 332712 & 204644 & 137786 & 0.0285 & 0.458  \\
\hline
\multirow{6}{*}{\rotatebox{90}{Cut}}&$n(\gamma)\geq 1$  & 316031 \!$\pm$\! 125& 192677 \!$\pm$\! 106& 129044 \!$\pm$\! 90.5& 0.0268 \!$\pm$\! 0.0407& 0.429 \!$\pm$\! 0.164 \\
\cline{2-7}
&$p_T(\gamma_1)> 150$~GeV  & 18576 \!$\pm$\! 132& 12146 \!$\pm$\! 106& 44616 \!$\pm$\! 173& 0.00831 \!$\pm$\! 0.07675& 0.142 \!$\pm$\!  0.313  \\
\cline{2-7}
&$\met > 150$~GeV  & 14681 \!$\pm$\! 118& 4287.3 \!$\pm$\! 64.8& 40297 \!$\pm$\! 168& 0.00743 \!$\pm$\! 0.07412& 0.127 \!$\pm$\! 0.303  \\
\cline{2-7}
&$n(j)\leq 1$, $|\eta(j)|< 4.5$  & 6819.7 \!$\pm$\! 81.7& 2388.2 \!$\pm$\! 48.6& 26564 \!$\pm$\! 146& 0.00475 \!$\pm$\! 0.06294& 0.0829 \!$\pm$\! 0.2605 \\
\cline{2-7}
&veto on $e$  & 6817.6 \!$\pm$\! 81.7& 1731.8 \!$\pm$\! 41.4& 26542 \!$\pm$\! 146& 0.00475 \!$\pm$\! 0.06292& 0.0828 \!$\pm$\! 0.2604  \\
\cline{2-7}
&veto on $\mu$  & 6817.6 \!$\pm$\! 81.7& 1132.5 \!$\pm$\! 33.6& 26536 \!$\pm$\! 146& 0.00475 \!$\pm$\! 0.06291& 0.0828 \!$\pm$\! 0.2604  \\
\cline{2-7}
&veto on $\tau$-jets  & 6479.8 \!$\pm$\! 79.7& 758 \!$\pm$\! 27.5& 25328 \!$\pm$\! 143& 0.00453 \!$\pm$\! 0.06172& 0.0789 \!$\pm$\! 0.2556 \\
\hline
\end{tabular}
\caption{The cut flow on signal and background events after requiring the parton level cuts $\slashed E_T>50$~GeV, $p_T(\gamma)>40$~GeV and $p_T(j_1)>30$~GeV for $M_{Z'}\simeq 2008$~GeV and $g_{B-L}=0.33$ (wide $Z'$ case) in the single-photon channel at $\sqrt s=14$~TeV with 
${\cal L}dt= 300$~fb$^{-1}$.}
\label{tab:monophoton2}
\end{center}
\end{table}

\section{$Z$-ISR signal}
\label{sec:monoZ}

Again, also in the case of the mono-$Z$ process, the Feynman diagrams which are relevant to the calculation are found
 in Fig.~\ref{fig:mono_feynman}.
For the mono-$Z$ signature, we generate events with the following parton level cuts: $\slashed E_T>80$~GeV, $p_T(\ell)>10$~GeV ($\ell = e,\;\mu$ ) and $p_T(j)>20$~GeV. The dominant irreducible background is $ZZ\to \ell^+ \ell^- \bar\nu\nu$ and the other background which is also irreducible is $WW\to \ell^+ \nu \ell^- \bar\nu $. The latter is controlled  after a cut in an invariant mass window centered on the $Z$ mass for two oppositely charged leptons, $m_{\ell^+\ell^-}\in [76,106]$~GeV \cite{Aad:2014vka}. The reducible backgrounds may have jets produced: $Z+\text{jets}$, $ZZ\to \bar{q} q \ell^+ \ell^-$ and $ZW\to \ell^+ \ell^- \bar{q}q$.  In addition, there are other reducible backgrounds with jet final states: (i) $t\bar{t} \to \ell^+ \nu b \ell^- \bar\nu\bar{b}$, which is reduced by rejecting events if they contain at least one jet with $p_T(j)>25$~GeV; (ii) $W+$jets, which is controlled by a large $\met$ cut. The last leptonic background is $ZW\to \ell\nu \ell^+ \ell^-$ .

\begin{figure}[t]
\includegraphics[width=8cm,height=8cm]{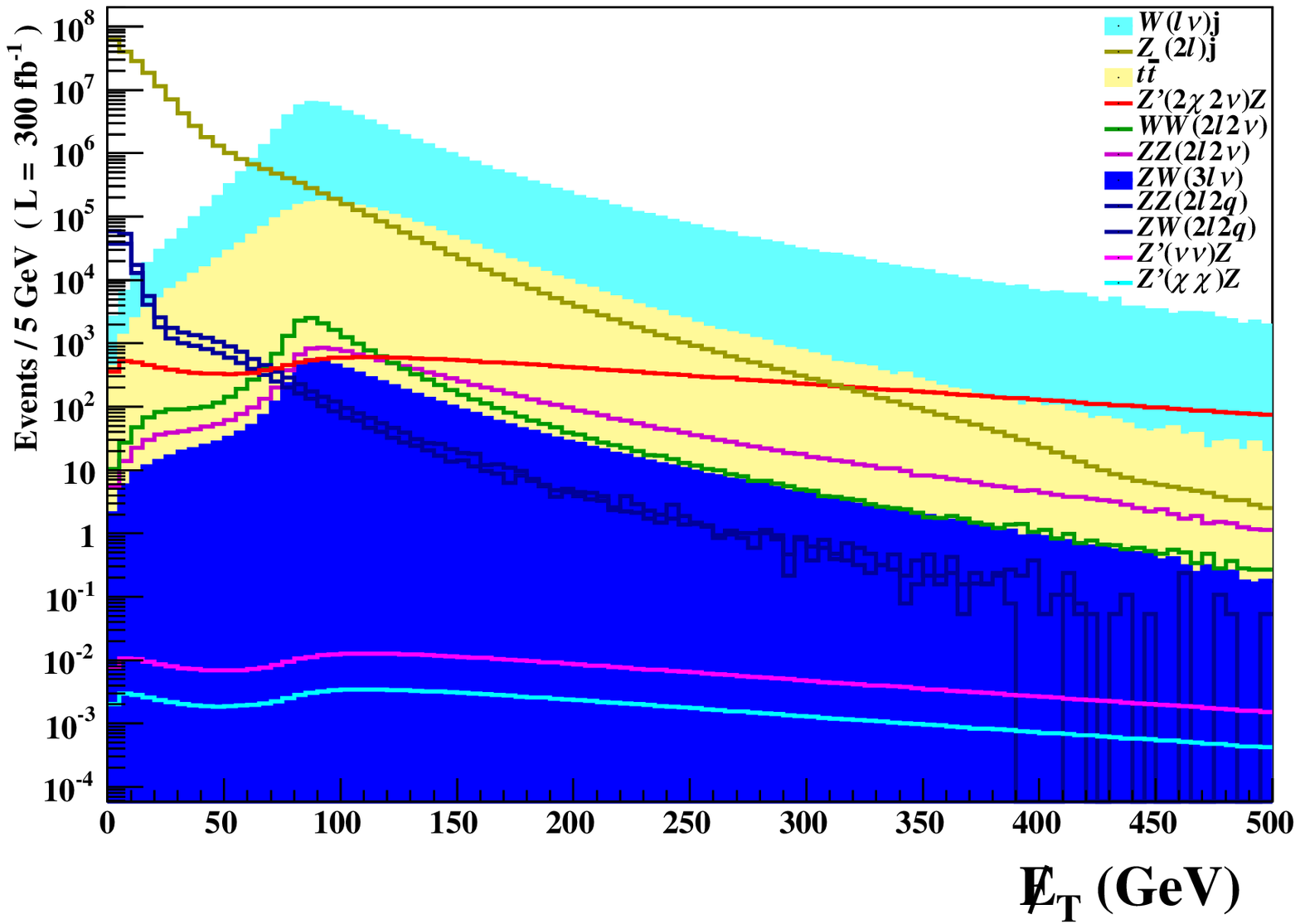}
\includegraphics[width=8cm,height=8cm]{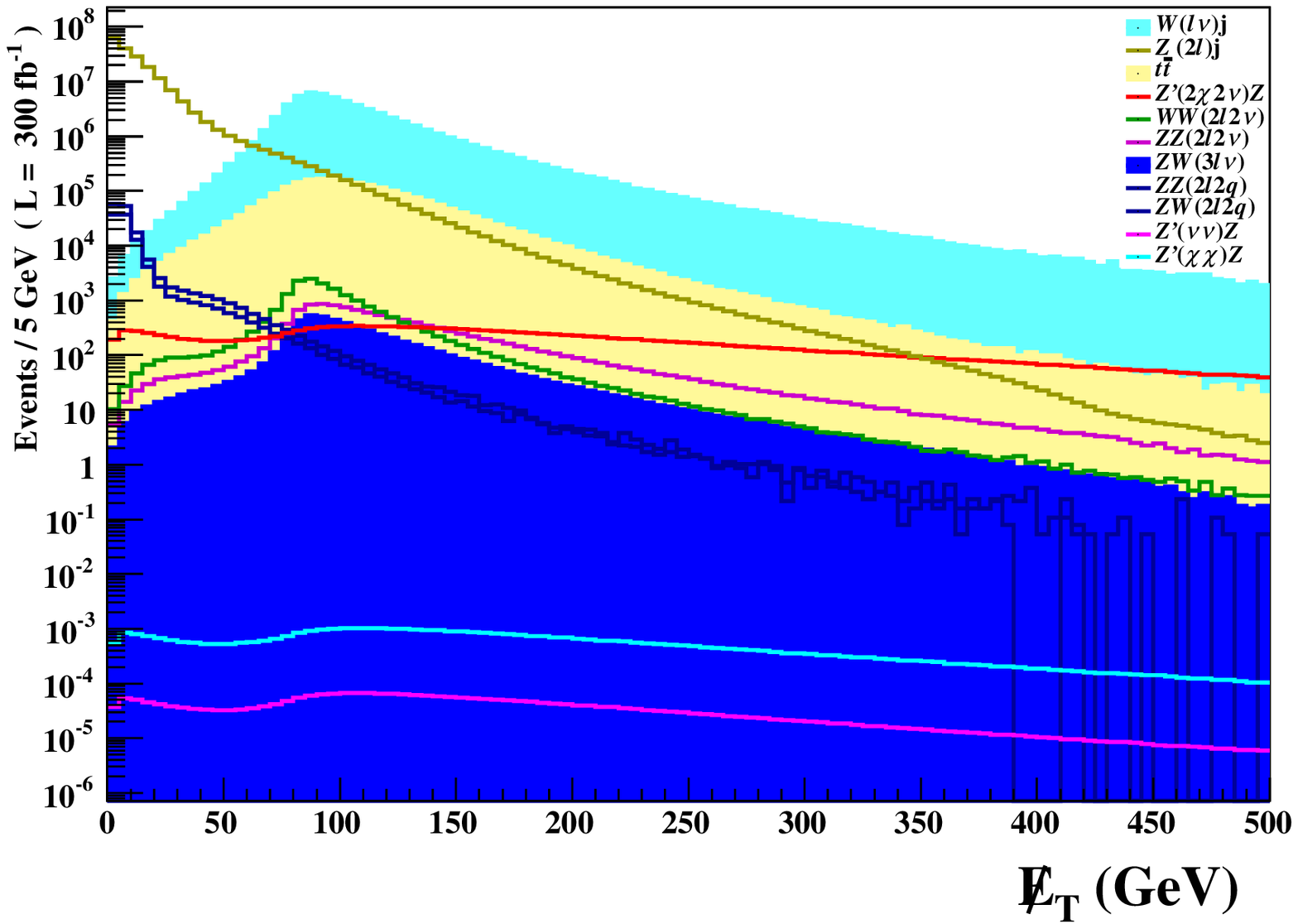}  
\caption{Number of events versus the missing transverse energy.
Both plots are presented before selection (i.e., detector level) cuts but after the parton level (i.e., MC generation) cuts 
$\slashed E_T>80$~GeV, $p_T(l)>10$~GeV and $p_T(j)>20$~GeV.
Rates are given at 14 TeV for an integrated luminosity of 300~fb$^{-1}$. (Left panel) $M_{Z'}\simeq 2448$~GeV and $g_{B-L}=0.4$ (narrow $Z'$ case). (Right panel)  $M_{Z'}\simeq 2008$~GeV and $g_{B-L}=0.33$ (wide $Z'$ case).
}
\label{fig5}
\end{figure}

\begin{figure}[t]
\begin{center}
\includegraphics[width=7.0cm,height=5.5cm]{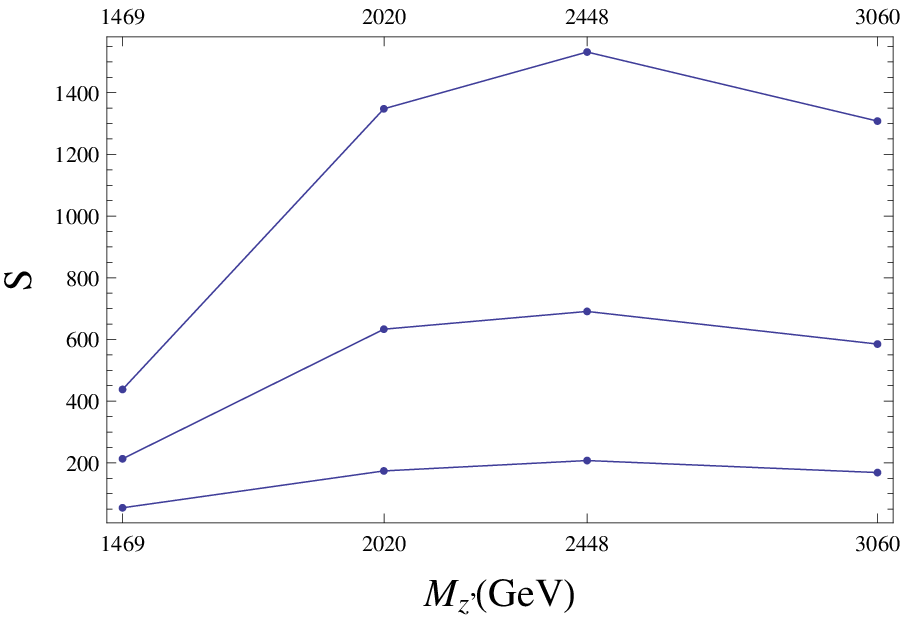}~\includegraphics[width=7.0cm,height=5.5cm]{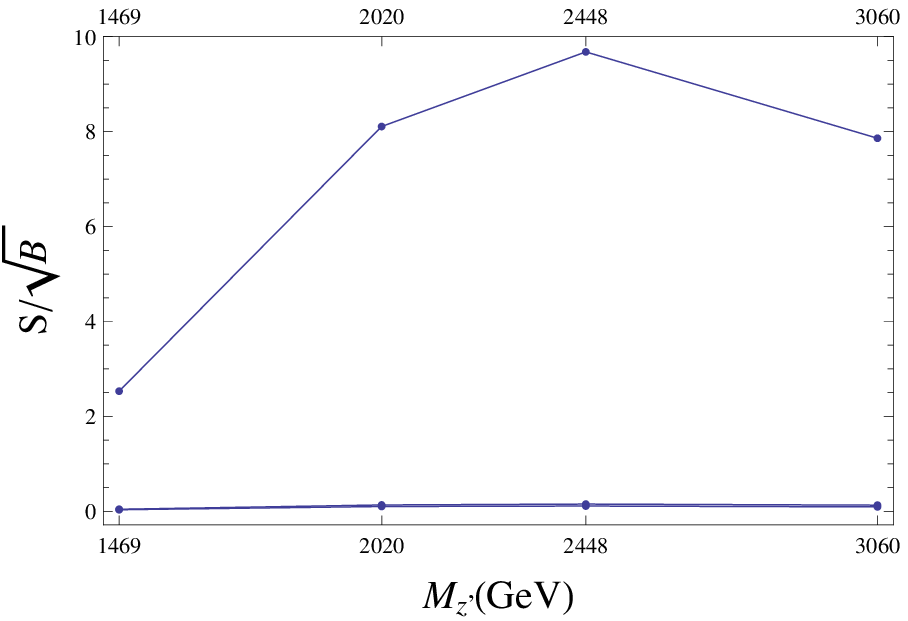}\\
\includegraphics[width=7.0cm,height=5.5cm]{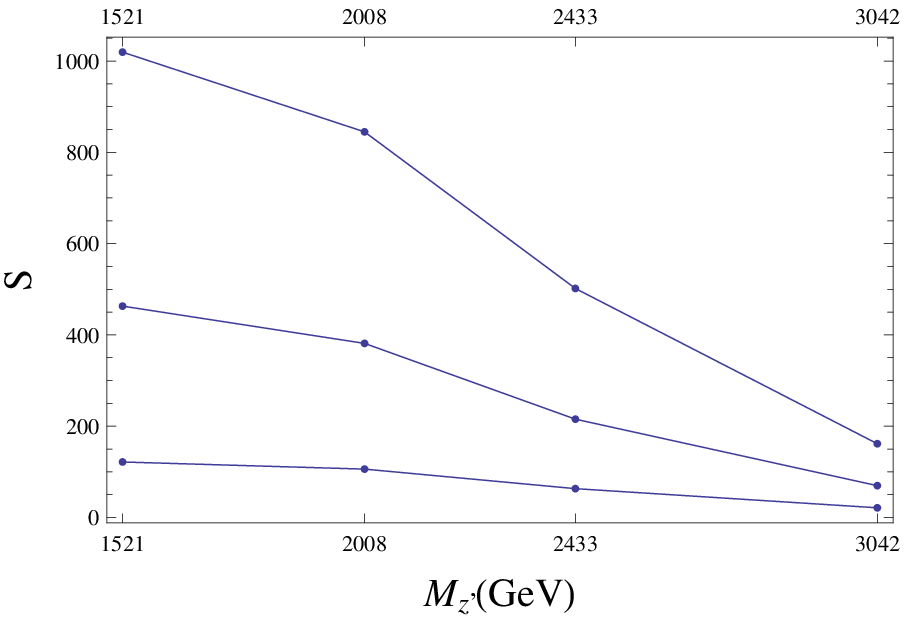}~\includegraphics[width=7.0cm,height=5.5cm]{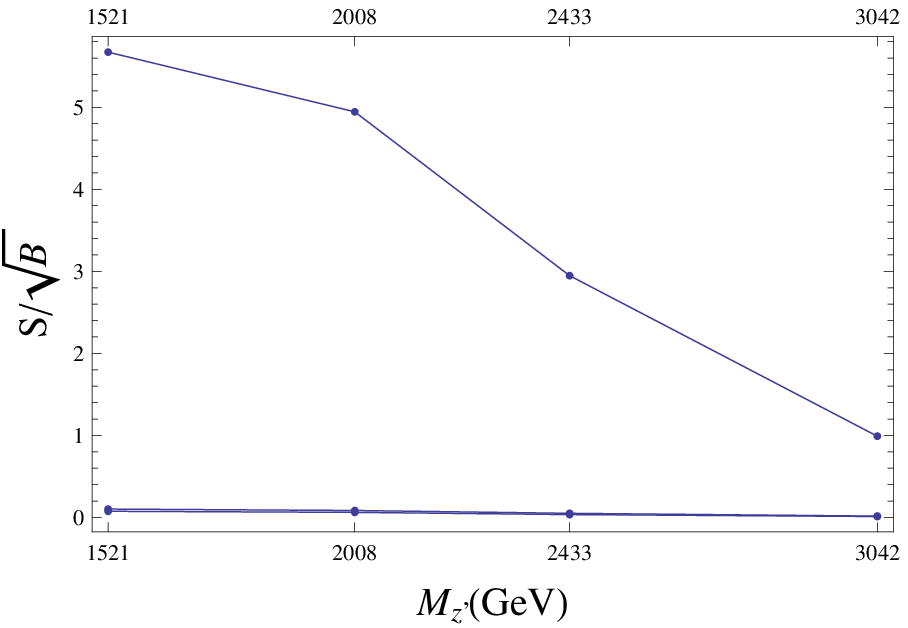}
\caption{(Top panel) Number of events from the sum of all signals $(S)$ versus $M_{Z'}$ and number of events from the sum of all signals divided by the square root of the total background $(S/\sqrt{B})$ versus $M_{Z'}$ for mono-$Z$ in the narrow $Z'$ case. (Bottom panel) Number of events from the sum of all signals $(S)$ versus $M_{Z'}$ and number of events from the sum of all signals divided by the square root of the total background $(S/\sqrt{B})$ versus $M_{Z'}$ for mono-$Z$ in the wide $Z'$ case. Rates are given at 14~TeV for an integrated luminosity of 300~fb$^{-1}$. } \label{monoZSBN}
\end{center}
\end{figure}

Motivated by  the plots in  Fig.~\ref{fig5}, where we show the signal and background distributions in $\met$, we adopt a
selection cut as follows: $\met > 250$~GeV. It is remarkable that after it the reducible backgrounds yield no event for the luminosity adopted while the irreducible ones are managable, see Tabs.~\ref{tab:monoZ1}--\ref{tab:monoZ2} \cite{Bell:2012rg}. Combine these results with those in Fig.~\ref{monoZSBN} to conclude that also the $Z$-ISR signal has some scope, certainly reduced
for discovery but potentially useful for diagnostics.

\begin{table}
\begin{center}
{\small\fontsize{7}{7}\selectfont{
\begin{tabular}{|@{\hspace{0.02cm}}c@{\hspace{0.02cm}}||@{\hspace{0.02cm}}c@{\hspace{0.02cm}}|@{\hspace{0.02cm}}c@{\hspace{0.02cm}}|@{\hspace{0.02cm}}c@{\hspace{0.02cm}}|@{\hspace{0.02cm}}c@{\hspace{0.02cm}}|@{\hspace{0.02cm}}c@{\hspace{0.02cm}}|@{\hspace{0.02cm}}c@{\hspace{0.02cm}}|@{\hspace{0.02cm}}c@{\hspace{0.02cm}}|@{\hspace{0.02cm}}c@{\hspace{0.02cm}}|@{\hspace{0.02cm}}c@{\hspace{0.02cm}}|}
\hline
\multicolumn{2}{|@{\hspace{0.02cm}}c@{\hspace{0.02cm}}|}{}&\multicolumn{5}{c|}{Backgrounds}&\multicolumn{3}{c|}{Signals}\\
\hline
\multicolumn{2}{|@{\hspace{0.02cm}}c@{\hspace{0.02cm}}|}{ Process} & $ZZ(2l2\nu)$ & $WW(2l2\nu)$ & $ZW(3l\nu)$ &$W(l\nu)j$& $t\bar{t}$  
&$Z'(2\tilde\chi \,2\nu)Z$&$Z'(\nu\bar{\nu})Z$&$Z'(\tilde\chi\tilde\chi)Z$ \\
\hline
\hline
\multicolumn{2}{|@{\hspace{0.02cm}}c@{\hspace{0.02cm}}|}{Before cuts} &12027 & 18966 & 5541 & 64980000 & 2377500 & 33900 & 0.703 & 0.191  \\
\hline
\multirow{3}{*}{\rotatebox{90}{Cut}}&(1)&9068.1 \!$\pm$\! 47.2 & 2726.2 \!$\pm$\! 48.3 & 4392.8 \!$\pm$\! 30.2& 521652 \!$\pm$\! 719 & 403272 \!$\pm$\! 578 & 1553.3 \!$\pm$\! 38.5& 0.0322 \!$\pm$\! 0.175& 0.0088 \!$\pm$\! 0.0914\\
\cline{2-10}
&(2)&6510.6 \!$\pm$\! 54.6& 2025.7 \!$\pm$\! 42.5& 2997.1 \!$\pm$\! 37.1& 193982 \!$\pm$\! 439& 12007 \!$\pm$\! 109& 696.2 \!$\pm$\! 26.1& 0.0145 \!$\pm$\! 0.119& 0.0039 \!$\pm$\! 0.0617  \\
\cline{2-10}
&(3)&229 \!$\pm$\! 15.0& 1.15 \!$\pm$\! 1.07& 49.63 \!$\pm$\! 7.01& 171 \!$\pm$\! 13.1& 8.76 \!$\pm$\! 2.96& 200.3 \!$\pm$\! 14.1& 0.0041 \!$\pm$\! 0.064& 0.0011 \!$\pm$\! 0.0334  \\
\hline
\end{tabular}}}
\caption{The cut flow on signal and background events after requiring the parton level cuts $\slashed E_T>80$~GeV, $p_T(l)>10$~GeV and $p_T(j)>20$~GeV for $M_{Z'}\simeq 2448$~GeV and $g_{B-L}=0.4$ (narrow $Z'$ case) in the $Z$-ISR channel at $\sqrt s=$ 14~TeV with ${\cal L}dt= 300$~fb$^{-1}$: (1) $ m_{ll}\in [76,106]$~GeV; (2) veto on $p_T(j)> 25$~GeV; (3) $\met > 250$~GeV.}
\label{tab:monoZ1}
\end{center}
\end{table}
\begin{table}
\begin{center}
{\small\fontsize{7}{7}\selectfont{
\begin{tabular}{|@{\hspace{0.02cm}}c@{\hspace{0.02cm}}||@{\hspace{0.02cm}}c@{\hspace{0.02cm}}|@{\hspace{0.02cm}}c@{\hspace{0.02cm}}|@{\hspace{0.02cm}}c@{\hspace{0.02cm}}|@{\hspace{0.02cm}}c@{\hspace{0.02cm}}|@{\hspace{0.02cm}}c@{\hspace{0.02cm}}|@{\hspace{0.02cm}}c@{\hspace{0.02cm}}|@{\hspace{0.02cm}}c@{\hspace{0.02cm}}|@{\hspace{0.02cm}}c@{\hspace{0.02cm}}|@{\hspace{0.02cm}}c@{\hspace{0.02cm}}|}
\hline
\multicolumn{2}{|@{\hspace{0.02cm}}c@{\hspace{0.02cm}}|}{}&\multicolumn{5}{c|}{Backgrounds}&\multicolumn{3}{c|}{Signals}\\
\hline
\multicolumn{2}{|@{\hspace{0.02cm}}c@{\hspace{0.02cm}}|}{ Process} & $ZZ(2l2\nu)$ & $WW(2l2\nu)$ & $ZW(3l\nu)$ &$W(l\nu)j$& $t\bar{t}$  
&$Z'(2\tilde\chi \,2\nu)Z$&$Z'(\nu\bar{\nu})Z$&$Z'(\tilde\chi\tilde\chi)Z$ \\
\hline
\hline
\multicolumn{2}{|@{\hspace{0.02cm}}c@{\hspace{0.02cm}}|}{Before cuts} &12027 & 18966 & 5541 & 64980000 & 2377500 & 18552 & 0.00325 & 0.0532  \\
\hline
\multirow{3}{*}{\rotatebox{90}{Cut}}&(1)&9068.1 \!$\pm$\! 47.2 & 2726.2 \!$\pm$\! 48.3 & 4392.8 \!$\pm$\! 30.2& 521652 \!$\pm$\! 719 & 403272 \!$\pm$\! 578 & 835.8 \!$\pm$\! 28.3& 0.00015 \!$\pm$\! 0.0119& 0.0025 \!$\pm$\! 0.0484\\
\cline{2-10}
&(2)&6510.6 \!$\pm$\! 54.6& 2025.7 \!$\pm$\! 42.5& 2997.1 \!$\pm$\! 37.1& 193982 \!$\pm$\! 439& 12007 \!$\pm$\! 109& 374.2 \!$\pm$\! 19.1& 0.00007 \!$\pm$\! 0.0082& 0.0011 \!$\pm$\! 0.0330  \\
\cline{2-10}
&(3)&229 \!$\pm$\! 15.0& 1.15 \!$\pm$\! 1.07& 49.63 \!$\pm$\! 7.01& 171 \!$\pm$\! 13.1& 8.76 \!$\pm$\! 2.96& 106.5 \!$\pm$\! 10.3& 0.00002 \!$\pm$\! 0.0041& 0.0003 \!$\pm$\! 0.0173  \\
\hline
\end{tabular}}}
\caption{The cut flow on signal and background events after requiring the parton level cuts $\slashed E_T>80$~GeV, $p_T(\ell)>10$~GeV and $p_T(j)>20$~GeV for $M_{Z'}\simeq 2008$~GeV and $g_{B-L}=0.33$ (wide $Z'$ case) in the $Z$-ISR channel at $\sqrt s=$ 14~TeV with ${\cal L}dt= 300$~fb$^{-1}$: (1) $ m_{ll}\in [76,106]$~GeV; (2) veto on $p_T(j)> 25$~GeV; (3) $\met > 250$~GeV.}
\label{tab:monoZ2}
\end{center}
\end{table}

\section{Summary and conclusions}
\label{sec:summa}

To recap our study, we have established the strong sensitivity that the LHC will have during the Run 2 stage 
with standard luminosity in probing invisible signals which may emerge in the BLSSM from $Z'$ decays in presence of an associated jet, photon or neutral weak boson. 

For all such signatures, upon enforcing newly developed selection procedures alongside standard triggers, 
we were in a position to access  significances well above the required 
$5\sigma$ discovery limit for all visible probes. This has been possible thanks to the fact that the BLSSM mediator of such invisible signals is a rather massive $Z'$ (with respect to the SM mediator of minimal SUSY, the $Z$ boson), with $M_{Z'}$ of
${\cal O}$(1 TeV), thereby transferring to its decay products large transverse momenta that can efficiently be exploited in all cases (mono-$j$, -$\gamma$ and -$Z$) for background reduction. Furthermore, of all the $Z'$ decay topologies considered here, the dominant one is via sneutrinos (above neutrinos and neutralinos), so that  extracting  these invisible signatures in the heavy missing transverse energy regime would not only signal the presence of a DM  induced channel within SUSY but also be a potential evidence of a theoretically well motivated non-minimal version of it, the BLSSM.
While the mono-jet sample would be used for discovery purposes owing to its high event rates and 
statistical significances, the mono-photon and -$Z$ data can be exploited to 
profile the underlying BLSSM signal, owing to the high level of experimental control achievable on $\gamma$ and especially
(leptonically decaying) $Z$ states. 

In this study,  we have proven the above points for a variety of BLSSM benchmark scenarios 
covering  both light ($\sim1.5$ TeV)  and heavy ($\sim3$ TeV)  as well as 
 narrow ($\sim100$ GeV)  and wide ($\sim800$ GeV) $Z'$ states, all compatible with the most recent experimental constraints from both EWPTs and LHC searches. Relevant (s)particle states ((s)neutrinos, charginos/neutralinos and
sleptons) were all taken below the TeV scale, hence accessible at the LHC, 
thereby offering alternative handles to establish the BLSSM.

Finally,
we have carried out our MC analysis at a rather sophisticated technical level, using multi-particle matrix element calculators,
parton shower plus hadronisation codes and detector emulation software, so as to believe that our results will withstand experimental tests.

\section*{Acknowledgments}
The work of W.A. and S.K. is partially supported by the STDF project 18448, the ICTP Grant AC-80 and the European Union FP7 ITN INVISIBLES (Marie Curie Actions, PITN-GA-2011-289442). J.F. and S.M. are financed in part through the NExT Institute. J.F. thanks the Galileo Galilei Institute in Florence, Italy, where part of this work was carried out, for hospitality. All authors are supported by the grant H2020-MSCA-RISE-2014 no. 645722 (NonMinimalHiggs). W.A. would like to thank M. Ashry, A. Ali and A. Moursy for fruitful discussions.

\end{document}